\documentclass[10pt,conference]{IEEEtran}
\IEEEoverridecommandlockouts
\usepackage{hyperref}
\usepackage{cite}
\usepackage{amsmath,amssymb,amsfonts}
\usepackage{algorithmic}
\usepackage{graphicx}
\usepackage{textcomp}
\usepackage{xcolor}
\def\BibTeX{{\rm B\kern-.05em{\sc i\kern-.025em b}\kern-.08em
    T\kern-.1667em\lower.7ex\hbox{E}\kern-.125emX}}

\newif\ifdraft
\ifdraft
\newcommand{\note}[1]{ {\textcolor{blue} { **: #1 }}}
\newcommand{\alnote}[1]{ {\textcolor{red} { ***Andre: #1 }}}
\newcommand{\prnote}[1]{ {\textcolor{green} { ***Philipp: #1 }}}
\newcommand{\jknote}[1]{ {\textcolor{blue} { ***Johannes: #1 }}}
\newcommand{\leonote}[1]{ {\textcolor{yellow} { ***Leo: #1 }}}
\newcommand{\jrfnote}[1]{ {\textcolor{orange} { ***Rudi: #1 }}}
\newcommand{\ieeenote}[2]{ {\textcolor{blue} { ***IEEE Reviewer #1: #2 }}}
\else
\newcommand{\note}[1]{}
\newcommand{\alnote}[1]{}
\newcommand{\prnote}[1]{}
\newcommand{\jknote}[1]{}
\newcommand{\leonote}[1]{}
\newcommand{\jrfnote}[1]{}
\newcommand{\ieeenote}[2]{}
\fi

\newcommand{\upp}{\vspace*{-0.5em}}

\usepackage{tikz}
\usepackage{mathtools}
\usepackage{listings}
\usepackage{url}
\usepackage{booktabs} 
\DeclareMathOperator*{\argmax}{arg\,max}
\DeclarePairedDelimiter\ceil{\lceil}{\rceil}

\usepackage[normalem]{ulem}
\newcommand\prsout{\bgroup\markoverwith{\textcolor{green}{\rule[0.5ex]{2pt}{0.7pt}}}\ULon}

\newcommand{\linebreakand}{%
  \end{@IEEEauthorhalign}
  \hfill\mbox{}\par
  \mbox{}\hfill\begin{@IEEEauthorhalign}
}

\begin{document}

\bstctlcite{IEEEexample:BSTcontrol}
\title{QUARK: A Framework for Quantum Computing Application Benchmarking
}

\author{\IEEEauthorblockN{Jernej Rudi Fin\v zgar\IEEEauthorrefmark{1}\IEEEauthorrefmark{2}\IEEEauthorrefmark{4},
Philipp Ross\IEEEauthorrefmark{1}\IEEEauthorrefmark{4}, 
Leonhard Hölscher\IEEEauthorrefmark{1}, Johannes Klepsch\IEEEauthorrefmark{1}, Andre Luckow\IEEEauthorrefmark{1}\IEEEauthorrefmark{3}}
\IEEEauthorblockA{\IEEEauthorrefmark{1}BMW Group, Munich Germany\\
                  \IEEEauthorrefmark{2}Technical University Munich, Germany\\
                  \IEEEauthorrefmark{3}Ludwig Maximilian University Munich, Germany\\
                  \IEEEauthorrefmark{4}{\footnotesize Authors contributed equally}\upp\upp\upp}}


\maketitle

\begin{abstract} 
 
Quantum computing (QC) is anticipated to provide a speedup over classical approaches for specific problems in optimization, simulation, and machine learning. With the advances in quantum computing toward practical applications, the need to analyze and compare different quantum solutions is increasing. While different low-level benchmarks exist, they often do not provide sufficient insights into real-world applica\-tion-level performance. We propose an application-centric benchmark method and the \emph{QUantum computing Application benchmaRK (QUARK)} framework to foster the investigation and creation of application benchmarks for QC.
This paper establishes three significant contributions:
(1) it makes a case for application-level benchmarks and provides an in-depth ``pen and paper'' benchmark formulation of two reference problems: \emph{robot path} and \emph{vehicle option optimization} from the industrial domain; (2) it proposes the open-source QUARK framework for designing, implementing, executing, and analyzing benchmarks; (3) it provides multiple reference implementations for these two reference problems based on different known, and where needed, extended, classical and quantum algorithmic approaches and analyzes their performance on different types of infrastructures. 
\end{abstract}

\begin{IEEEkeywords}
quantum computing, benchmark, optimization
\end{IEEEkeywords}

\section{Introduction}
\emph{Motivation:} Quantum computing (QC) is transitioning from research to industrialization. It promises to improve optimization, machine learning, and simulation problems significantly, overcoming the limitations of existing high-performance computing systems~\cite{mck_2021}. Applications for these problem domains can be found in academia and industry~\cite{qutac_epj}. For example, the automotive industry's complex design, manufacturing, logistics, and financial challenges are promising candidates for quantum-based optimization and machine learning approaches. Quantum chemistry simulations promise to enhance the material research process, e.\,g., for battery cell chemistry.


Impressive progress has been made, as visible, e.\,g., in several quantum advantage demonstrations~\cite{Arute2019,xanadu}. However, it is currently unclear what hardware technology and algorithms will deliver a practical quantum advantage, i.\,e., a quantum system that provides better solution quality, time-to-solution, energy usage, or cost than a classical system. 
The evaluation of quantum systems is becoming increasingly important to assess scientific and technical progress addressing the needs of end users, funding agencies, and investors. Benchmarks are critical for this purpose and will help to guide application, algorithm, and hardware development, and build communities~\cite{1201189,bcg2022}. 


\emph{State-of-the-art and limitations:} 
Current quantum computing benchmarks often focus on low-level hardware performance, targeting hardware providers~\cite{PhysRevA.100.032328,wack2021quality} and providing valuable metrics to assess technological maturity and roadmaps. Unfortunately, the results often do not translate to real-world application performance.
Higher-level benchmarks, e.\,g., Lubinski~\cite{lubinski2021applicationoriented} and Martiel~\cite{martiel2021benchmarking}, consider a set of algorithms and circuits. While these approaches provide important insights, they do not investigate end-to-end application performance and thus, foster holistic advances on all levels required for real-world applications.



\emph{Key insights, contributions, and artifacts:} In this work, we make three significant contributions: 

(1) We propose an application-centric approach for developing benchmarks. By using a ``pencil and paper'' approach (as popularized by the NAS parallel benchmark (NPB)~\cite{Bailey2011}), we allow for multiple problem formulations, e.\,g., quantum annealing, gate-based, hybrid and classical formulations. Considering the maturity of quantum hardware and programming systems, we think this approach is best-suited, 
facilitating innovations and optimizations on all levels, e.\,g., hardware, control system, operating system and middleware, algorithm, and application level. Specifically, we provide a formulation of two reference problems from the industrial domain, \emph{robot path} and \emph{vehicle option} optimization (see section~\ref{sec:applications});

(2) We introduce the open-source \emph{QUantum computing Application benchmaRK} (QUARK)~\cite{quarkGithub} framework for designing, implementing, executing, and analyzing benchmarks (see section~\ref{sec:quark}). QUARK addresses critical  requirements of application benchmarks, such as the need to abstract realistic workloads and datasets into benchmarks, support multiple implementations, and reproducibly capture all results;

(3) We demonstrate QUARK's capabilities by implementing benchmarks for the two reference problems (see section~\ref{sec:results}). For these problems, we develop and characterize several classical and quantum algorithms (e.\,g., a novel QUBO formulation of the partial MAX-SAT problem) and benchmark these on different infrastructures (e.\,g., D-Wave and simulation). 

\emph{Limitations:} It is challenging to develop representative application benchmarks for quantum computing, as it is unclear which algorithm, qubit modality, and hardware will deliver a quantum advantage. As current quantum systems provide no real practical advantage, the utility of application-level benchmarks is still limited.  Further, transferring benchmark results to other applications is often challenging. Finally, in its present form QUARK is limited to optimization problems and thus does not cover all quantum application domains.

\section{Background and Related Work}
\label{sec:related_work}
\subsection{Quantum Computing Infrastructure}

Various hardware realizations of quantum computers have been proposed and are in development.
These quantum hardware systems typically possess different characteristics, e.\,g., gate fidelities, coherence times, and clock speeds. Currently, superconducting and ion-trapped qubits are the most widely used modalities. Both modalities are available from different vendors, e.\,g., superconducting systems from IBM~\cite{ibm}, Google~\cite{google}, and Rigetti~\cite{rigetti} and ion-trap-based systems from IonQ~\cite{ionq} and Honeywell~\cite{honeywell}. Non-gate-based systems for quantum annealing from D-Wave are also broadly available on the D-Wave~\cite{dwave} and AWS clouds.
Finally, approaches such as neutral atom~\cite{lukin-neutral}, and topological quantum computation~\cite{topological-qc} could become prominent in the future.

Additionally, classical simulation of quantum systems is crucial for designing quantum algorithms and verifying results obtained on quantum devices. Hence, it is necessary to understand the trade-offs and scales of different simulation approaches (see~\cite{qc_sim_list} for an overview).
\subsection{Benchmarks}

%
Benchmarks are standardized workloads, i.\,e., sets of inputs (program and data), that are used to compare computer systems~\cite{ferrari1978computer, Jain:1991:PerformanceAnalysis} and have been instrumental in many areas of computer science and engineering. 
In general, two types of benchmarks exist: (i) specification-based benchmarks that provide a ``pen and paper'' description of a problem; and (ii) reference implementations. Both approaches have trade-offs: specification-based benchmarks are flexible, allowing for innovation. Results of these benchmarks are, however, difficult to compare. Reference implementations limit the design space and allow for more controlled yet expensive experiments. 

Benchmarks arise on different levels: System-level benchmarks focus on the lower-level hardware and system aspects (e.\,g., gate fidelity) and, thus, are difficult to map to application performance. Algorithmic-level benchmarks evaluate specific, significant subroutines. Application-level benchmarks are more holistic and consider the entire stack comprising hardware, operating system, middleware, classical resources, and the interplay between the individual components.
Nevertheless, transferring insights and results between different applications is often difficult as only a narrow set of these interactions can be covered. Table~\ref{tab:benchmarks} summarizes classical and quantum benchmarks for different levels.

\begin{table}
 \caption{Important benchmarks for the different layers: From system-level to application-level benchmarks. \label{tab:benchmarks} }
\resizebox{0.97\columnwidth}{!}{%
\begin{tabular}{p{1.45cm}p{4.2cm}p{4.2cm}}
\toprule
Level & Classical & Quantum \\ \midrule
Application           &ImageNet~\cite{imagenet}, Glue~\cite{glue}, MLPerf~\cite{2019arXiv191001500M}, Chook~\cite{perera2021chook}           &QScore~\cite{martiel2021benchmarking}, QED-C~\cite{lubinski2021applicationoriented}, SupermarQ~\cite{supermarq},  Fermi-Hubbard Model~\cite{gard2021classically,dallairedemers2020application}\\
Algorithm             &Linpack~\cite{Dongarra02thelinpack}, NPB~\cite{Bailey2011},  SPEC~\cite{spec}, TSPLib95~\cite{RePEc:inm:orijoc:v:3:y:1991:i:4:p:376-384}, 
                            SAT competition~\cite{sat}         
                                       &VQE~\cite{mccaskey2019quantum}, QAOA~\cite{Willsch}, Annealing~\cite{Katzgraber_PhysRevApplied.12.014004,yarkoni2021multicar} \\
                                       System                &SPEC HPC, ACCEL~\cite{accel}, MPI~\cite{mpi}, OMP~\cite{spec}                              &QV~\cite{PhysRevA.100.032328}, Volumetric benchmarking~\cite{BlumeKohout2020volumetricframework}, randomized gate benchmarking~\cite{PhysRevA.100.032328}, Arline compiler benchmark~\cite{arline}, CLOPS~\cite{wack2021quality}, QASMBench~\cite{QASMBench}
                                               \\ \bottomrule
\end{tabular}
}\upp\upp
\end{table}

\subsubsection{Classical Benchmarks}

Important benchmarks relevant to quantum computing have emerged in HPC and in many application domains. For example, the High-Performance Linpack (HPL)~\cite{Dongarra02thelinpack} 
is used to create the Top500 supercomputing list.\jrfnote{Cut the NPB stuff, since it's already mentioned in the intro?} The NPB~\cite{Bailey2011} originated in the domain of aerodynamics simulations and is a ``paper and pencil'' benchmark, comprising five parallel kernels, and three application benchmarks (e.\,g., LU matrix decomposition).

Various benchmarks have been proposed and advanced with the emergence of data and machine learning workloads and applications. Application-centric benchmarks, such as ImageNet~\cite{imagenet} for computer vision and Glue~\cite{glue} for natural language processing, were instrumental in advancing the state of machine learning, by providing labeled, standardized datasets that enabled comparisons.

There exist several benchmarks for common optimization tasks, such as the Boolean satisfiability problem (SAT)~\cite{sat}, scheduling~\cite{shift_scheduling, clusterdata:Wilkes2020a}, and the traveling salesperson problem (TSP)~\cite{RePEc:inm:orijoc:v:3:y:1991:i:4:p:376-384}.



\subsubsection{Quantum System Benchmarks}


Quantum system benchmarks focus on low-level aspects of quantum devices. 
One prominent example is the quantum volume (QV) benchmark~\cite{PhysRevA.100.032328}, defined as the largest executable circuit with equal width (number of qubits) and depth (number of circuit layers). 
Thus, the QV provides valuable information (e.\,g., gate fidelities, coherence times) needed to assess quantum hardware quality and to validate roadmaps.

Blume-Kohout et al.~\cite{BlumeKohout2020volumetricframework} extend the QV beyond square circuits, allowing rectangular circuits with different numbers of qubits and layers. Further, the authors propose, in addition to randomized circuits used by QV, the use of other circuit types, e.\,g., Grover iterations and Hamiltonian simulations, and additional quality metrics.

The circuit layer operations per second (CLOPS) metric focuses on the execution speed~\cite{wack2021quality}. The benchmark is based on parametrized circuits, i.\,e., circuits which are static and are configured with parameters at runtime. Parameterized circuits are used in quantum algorithms for machine learning, optimization, and chemistry, particularly in the NISQ-era. The metric considers the circuit execution time, including, e.\,g., preparation overheads.

QASMBench~\cite{QASMBench} is a benchmark suite providing different small to large-scale quantum circuits with an emphasis on evaluating these circuits on quantum hardware. While the benchmark includes application-relevant circuits for  (e.\,g., the quantum approximate optimization algorithm (QAOA)), the authors focus on evaluating quantum hardware performance using these circuits (e.\,g., gate fidelity).

\subsubsection{Quantum Application Benchmarks}

We differentiate between characterizations, i.\,e., activities that focus on describing and understanding systems, and benchmarks, i.\,e., standardized workloads that allow a comparison between systems.

\emph{Characterizations:} 
D-Wave devices have been thoroughly investigated regarding their performance, tunability, and limitations for different applications from science to finance and industry.
Grant et al.~\cite{osti_1756458} utilize a portfolio optimization use case to analyze the effects of different control parameters of quantum annealers. 
In particular, they monitored how the solution quality changes with different embeddings, annealing times, and spin reversal routines. 
Perdomo et al.~\cite{Katzgraber_PhysRevApplied.12.014004} investigate the combinational circuit fault diagnosis (CCFD) industrial optimization problem, focusing on the scalability of annealing approaches.


Various characterizations of gate-based systems and applications exist. Willsch et al.~\cite{Willsch} investigate the performance of QAOA and annealing and their ability to discover the optimal solution for artificial Max-Cut and 2-SAT problems. Performance aspects, e.\,g., the time-to-solution, are not investigated.



\emph{Benchmarks:} 
While previous examples focus on specific application scenarios, Mills et\,al.~\cite{Mills_2021} emphasize the need for more holistic benchmarks. To this end, the authors propose three circuit designs: shallow, square, and deep circuits. The proposed approach is similar to the volumetric benchmark approach proposed in~\cite{BlumeKohout2020volumetricframework}. While these circuit types can be mapped to more concrete applications on a high level, it is difficult to predict performance on concrete applications (e.\,g., for specific problem types and sizes).  


Martiel et al.~\cite{martiel2021benchmarking} propose an application-centric optimization benchmark called Q-Score. The Q-Score is based on performing the Max-Cut algorithm using QAOA on different sizes of standardized Erdös-Renyi graphs. As Q-Score only encapsulates a single problem, its practical value is limited.

Lubinski et\,al.~\cite{lubinski2021applicationoriented} propose application-oriented benchmarks to assess gate-based quantum systems using a volumetric framework. Currently, the framework comprises 11 different algorithms.
While most of these algorithms provide important building blocks for quantum applications, the analysis is not conducted in the context of industry applications. 
Important application domains, such as optimization and machine learning, are not addressed. The framework relies on a normalized fidelity metric, comparing the output distributions of the optimal solution and experiment.

Tomesh et\,al.~\cite{supermarq} propose the SupermarQ benchmark suite comprising eight benchmark kernels (e.\,g., GHZ, Flip Code, and QAOA) for gate-based devices. It focuses on synthetic problems, e.\,g., QAOA-based Max-Cut optimization on a random graph, making it difficult to generalize results to real-world problems.

\emph{Discussion:}
Most approaches focus on specific applications and systems, investigating different configurations to improve understanding. Further, they often rely on generic quality metrics, which are difficult to map to real-world application performance. Finally, they often lack an end-to-end perspective and ignore hidden costs, e.\,g., the time required to move data between classical and quantum interfaces. 
The current state reflects the maturity of the quantum ecosystem, which is yet to deliver a practical advantage. The standardization of metrics, datasets, benchmarking methods, and reproducibility will become increasingly important considering the rapid progress toward real-world applications.  

\section{Applications and Workloads}
\label{sec:applications}
\ieeenote{3}{One opportunity for improvement is in the justification of the selection of the two representative applications. Why are these industrially relevant? The justification presented in the paper is cursory and very generic, and lacks any connection to specific industry problems or value propositions. Additionally, little to no explanation is provided as to why these are relevant for quantum benchmarks as well. This is an area in which more details would be greatly appreciated and would elevate the work.}


Solving optimization problems has been a major driving force behind the development of quantum computing, as even marginal improvements over existing methods can lead to significant economic impact.
This section describes and rigorously formulates two representative industrial applications from the optimization domain: \emph{robot path planning} and \emph{vehicle options planning}. The results of both benchmark problems provide immediate utility for assessing a proposed solution's feasibility and business impact. Further, various ways exist to scale and adapt the problems to the evolving technological landscape.

\ieeenote{2}{My only very minor criticism is that the problems were stripped of some of their complexity/size to allow for the benchmarks to be properly ran. While this is useful and should be done, I think maintaining the problem in its full, industrially-relevant form as a 'challenge' problem would add a lot of value.}\jrfnote{I do not completely understand what they mean by this -- I think the problem is described quite generally in both cases?}



\subsection{Robot Path Planning}

\paragraph*{Application}
Robots are a crucial enabler for automation in industrial manufacturing, driving quality, efficiency, and scale improvements.  However, the deployment of robot systems comprising software and hardware is challenging. One particular example is planning paths for complex multi-robot systems~\cite{9108020}. Robots must follow a pre-defined path to execute multiple tasks in such systems.

An example is the polyvinyl chloride (PVC) sealing process, in which spaces on the vehicle body, e.\,g., between joint sheets, are sealed using PVC, a thermoplastic material, to increase waterproofness and prevent corrosion. 
The real-world system is highly complex -- for example, each robot has multiple tools and configuration settings, like the number and type of nozzles used by each robot.
Multiple robots (up to four) work in parallel during this process. Thus, spatial constraints to avoid collisions must be enforced. The objective is to find the shortest valid path that fulfills the following requirements: (1) all seams need to be sealed; (2) the robot must always start and end at a particular home position; and (3) no collision between the different robots may occur.

Due to the limitations of current quantum computing hardware, we make several simplifications. First, we only consider single robot systems, i.\,e., no collisions must be avoided. Second, we simplify the dataset and aggregate data across some dimensions, e.\,g., different available tools and configuration parameters. Third, we only consider two different tools and configuration settings. Further, we decrease the problem size to allow execution on current quantum hardware. For this purpose, we remove seams from the real-world problem graph deterministically to ensure reproducibility.  





\paragraph*{Problem Class}
The problem is a variant of the NP-hard TSP. It is specified using a weighted graph, encoding the distances between all possible node pairs. The goal is to find a combination of nodes representing the shortest path and, thus, the shortest time. 

While robot path planning shares some similarities with the TSP, there are some key differences:
(1) There are two nodes per seam, but only one of these nodes needs to be visited to seal that seam; (2) there are numerous tools and configuration settings in which a node can be visited; (3) the costs from one node to the other with a specific tool/configuration setting are not symmetric; and (4) the graph is not fully connected as not all moves are possible.





\paragraph*{Mathematical Model}
\label{sec:applications_pvc_math}


We define  $x_{s n c t }^{(i)}$ as a binary variable, which we set to $1$ if the robot is at the node $(s, n, c, t)$ at time-step $i$, where $s$ denotes the seam number, $n$ the node number, $c$ the configuration and $t$ the tool setting. Overall there are $N_{\mathrm{seams}} +1$ time-steps as we need to visit all seams plus the special home position for a path to be valid.
The cost function comprises the following components:
\begin{subequations}
\begin{align}
    \label{eq:pvc_qubo_dis_1}
    f_{\mathrm{dist.}}(\mathbf{x}) &=  \sum_{i=1}^{N_{\mathrm {seams }}+1} \sum_{(s, n, c, t)} \sum_{\left(s^{\prime}, n^{\prime}, c^{\prime}, t^{\prime}\right)} d_{s n c t }^{s^{\prime} n^{\prime} c^{\prime} t^{\prime}} x_{s n c t }^{(i)} x_{s^{\prime} n^{\prime} c^{\prime} t^{\prime}}^{(i+1)},  \\ 
    \label{eq:pvc_qubo_dis_2}
    f_{\mathrm{time}}(\mathbf{x}) &=\sum_{i=1}^{N_{\mathrm{seams }}+1}\left[\sum_{(s, n , c, t)} x_{s n c t}^{(i)}-1\right]^{2},
    \\
    \label{eq:pvc_qubo_dis_3}
    f_{\mathrm{comp.}}(\mathbf{x}) &= \sum_{s=1}^{N_{\mathrm{seams }}+1}\left[\sum_{i=1}^{N_{\mathrm{seams }}+1}\sum_{(n, c, t)} x_{s n c t}^{(i)}-1\right]^{2},
\end{align}
\end{subequations}
where we have collected all $x_{s n c t }^{(i)}$ into a vector $\mathbf{x}$.
For simplicity, the home position is included in $(s, n, c, t)$.
The total distance covered by the robot is $f_{\mathrm{dist.}}$, with $d_{s n c t }^{s^{\prime} n^{\prime} c^{\prime} t^{\prime}}$, representing the distance between $x_{s n c t }$ and $x_{s^{\prime} n^{\prime} c^{\prime} t^{\prime}}$. 
Additionally, we defined two constraint terms: $f_{\mathrm{time}}$ and $f_{\mathrm{comp.}}$. The constraint term $f_{\mathrm{time}}$ ensures that only a single node is visited per time-step, while $f_{\mathrm{comp.}}$ ensures that every task is performed exactly once, i.\,e., every seam is sealed and the home position is visited.

 

The total cost function is given by
\begin{align}
\label{eq:pvc_qubo_total}
f(\mathbf{x}) =  f_{\mathrm{dist.}}(\mathbf{x}) + \lambda\left[f_{\mathrm{comp.}}(\mathbf{x}) + f_{\mathrm{time}}(\mathbf{x})\right],
\end{align}
where $\lambda$ is the Lagrange parameter determining the magnitude of the constraint terms.
The resulting QUBO instance can be optimized using quantum approaches such as quantum annealing, QAOA, or classical algorithms.
In a post-processing step, we reorder the solution so that the robot starts at the home position.

Using 
\begin{subequations}
\begin{align}
\label{eq:pvc_qubit_formula}
\begin{split}
N_{\mathrm{qubits}} 
={} &
( 2 \text{ \tiny(nodes per seam)}
\cdot
N_{\mathrm{seams}}
+
1 \text{ \tiny(home position)})
\\
& \cdot
N_{\mathrm{configs}}
\cdot
N_{\mathrm{tools}}
\cdot
N_{\mathrm{time-steps}},
\end{split}
\\
\label{eq:pvc_qubit_formula_time-steps}
N_{\mathrm{time-steps}}
={}
& N_{\mathrm{seams}}
+
1 \text{ \tiny(home position)},
\end{align}
\end{subequations}
we can compute the number of qubits $N_{\mathrm{qubits}}$ needed to encode the optimization objective Eq.~\eqref{eq:pvc_qubo_total} on a quantum device (see Table~\ref{Tab:qubitspvc}).

\begin{table}
    \centering
    \caption{Resource estimation for the robot path optimization, displaying the number of qubits required for problem instances of increasing complexity.}
    \label{Tab:qubitspvc}
    \begin{tabular}{rrrrr}
        \toprule
        $N_{\mathrm{seams}}$ &  $N_{\mathrm{tools}}$ &  $N_{\mathrm{configs}}$ &  $N_{\mathrm{time-steps}}$ &  $N_{\mathrm{qubits}}$ \\
        \midrule
             1 &      2 &       2 &          2 &             24 \\
             2 &      2 &       2 &          3 &             60 \\
            \vdots &      \vdots &       \vdots &         \vdots &          \vdots \\
            70 &      4 &       4 &         71 &         160176 \\
        \bottomrule
    \end{tabular}
\upp\upp\upp
\end{table}


\subsection{Vehicle Options}
\label{sec:vehicle-options}



\paragraph*{Application}
Before a new vehicle model can be deployed for production, several tests have to be carried out on pre-series vehicles to ensure the feasibility and  the functionality of specific component configurations.
The manufacturer wants to save resources and produce as few pre-series vehicles as possible while still performing all desired tests. 
Further, not all feature configurations can realistically be implemented in all vehicles, leading to constraints that the produced vehicles must satisfy.
\paragraph*{Problem Class}
The vehicle options optimization problem belongs to the family of SAT problems, which are hard to solve.
SAT problems ask whether a configuration of Boolean variables exists, such that a given Boolean formula evaluates to $1$.\footnote{We use $\mathrm{False}\equiv 0$ and $\mathrm{True}\equiv 1$ throughout this manuscript.}
SAT problems are NP-complete and not only lie at the center of contemporary theoretical computer science research but also appear in a wide range of fields, such as artificial intelligence~\cite{sat-ai}, circuit design~\cite{sat-circuit-design}, and computational biology~\cite{sat-bioinformatics}.
Additionally, an extension of SAT problems to maximum satisfiability (MAX-SAT) is frequently used to search for the configuration that maximizes the number of satisfied clauses.
Due to its theoretical importance and applicability, the study of (MAX-)SAT is an active area of research~\cite{sat,satproceedings2021}.

\paragraph*{Mathematical Model}
Consider the set of $N_{\mathrm{v}}$ test vehicles $\left\{\mathbf{v}^{(1)},...,\mathbf{v}^{(N_{\mathrm{v}})}\right\}$, where each vehicle is exactly defined by its configuration of $N_{\mathrm{f}}$ features.
That is, for each $i$, $\mathbf{v}^{(i)}\in\left\{0,1\right\}^{N_\mathrm{f}}$ is a binary vector of dimension $N_{\mathrm{f}}$, where its $j$th component $v^{(i)}_j$ encodes the information whether feature $j$ is absent ($v^{(i)}_j=0$) or present ($v^{(i)}_j=1$) in this particular vehicle.

In a realistic setting, not all $2^{N_\mathrm{f}}$ possible configurations are feasible (e.\,g., a vehicle cannot simultaneously have a V4 and V8 engine), 
leading to the introduction of 
$N_{\mathrm{h}}$ constraints $\phi_k$.
Each constraint can be specified as a Boolean expression involving some subset of features. 
For example, the condition that vehicle $i$ must contain at least one of the features 1 or 2, and not include feature 3 can be formulated as follows:
$$
    \phi_{\mathrm{example}}\left(\mathbf{v}^{(i)}\right)
    =
    \left(
    v^{(i)}_1\lor v^{(i)}_2
    \right)
    \land
    \overline{v}^{(i)}_3.
$$

Since all of the $n$ vehicles have to satisfy each of the $p$ constraints, this means that we require
\begin{equation}    
    \bigwedge_{j=1}^{N_\mathrm{h}}
    \bigwedge_{i=1}^{N_\mathrm{v}}
    \phi_j\left(\mathbf{v}^{(i)}\right) 
    =
    1.
\end{equation}

Additionally, we want to perform ${N_\mathrm{s}}$ different tests on the vehicles. 
We model this by introducing a collection of ${N_\mathrm{f}}$ test requirements $\theta_i$ -- 
we demand each of the $\theta_i$ to be satisfied by at least one of the ${N_\mathrm{v}}$ vehicles: 
\begin{equation}
    \bigwedge_{k=1}^{N_\mathrm{s}}
    \bigvee_{i=1}^{N_\mathrm{v}}
    \theta_k \left(\mathbf{v}^{(i)}\right)
    =
    1.
\end{equation}
Combining the buildability constraints and the test requirements, we can state the full mathematical formulation of the vehicle options problem as:
\begin{equation}
    \left[
    \bigwedge_{j=1}^{N_\mathrm{h}}
    \bigwedge_{i=1}^{N_\mathrm{v}}
    \phi_j \left(\mathbf{v}^{(i)}\right)
    \right]
    \land
    \left[
    \bigwedge_{k=1}^{N_\mathrm{s}}
    \bigvee_{i=1}^{N_\mathrm{v}}
    \theta_k \left(\mathbf{v}^{(i)}\right)
    \right]
    =
    1.
\end{equation}

In practice, a related question is asked: given that a certain quantity of vehicles can be produced, what is the configuration of features of the produced vehicles that maximizes the number of tests that can be performed on them?
Due to the limited capabilities of current quantum devices, we limit ourselves to finding the optimal configuration of features for a single vehicle.
This approach can be interpreted as a single step of the optimization procedure for multiple vehicles. After one finds the vehicle that satisfies the most tests, the tests that have been satisfied can be removed from consideration. The next chosen vehicle is chosen by maximizing the number of the remaining tests.

Thus, the optimal configuration is defined as:
\begin{equation}
    \mathbf{v}^*
    =
    \argmax_{
    \mathbf{v}\in\Phi
    }\left(
    \sum_{k=1}^{N_\mathrm{s}} \theta_k(\mathbf{v})
    \right),
\end{equation}
where $\Phi=\left\{\mathbf{v}\mid\bigwedge_{k=1}^{N_\mathrm{h}}\phi_k\left(\mathbf{v}\right)=1\right\}$ is the set of the configurations that satisfy all buildability constraints\footnote{In principle, a set of weights $\{w_k\}$ could be used to modify the objective function to $\sum_{k=1}^{N_\mathrm{s}} w_k\psi_k(\mathbf{v})$, yielding a weighted (partial) MAX-SAT instance. This  corresponds to prioritizing some tests.}.
This problem formulation is an instance of MAX-SAT,\alnote{consider removing that footnote: first the reader is interrupted in the flow and the information of the alternative name if important could be put in parentheses} \footnote{In the literature, Partial MAX-SAT is sometimes used to describe such problems.} with the buildability constraints and test requirements corresponding to \emph{hard} and \emph{soft constraints} from the MAX-SAT literature~\cite{softhardcons}.

For simplicity, we limit ourselves to MAX-3SAT (i.\,e., MAX-SAT where all clauses are length 3) instances in conjunctive normal form (CNF), since any MAX-SAT instance can efficiently be brought into this form~\cite{skiena2008the, tseitin}.




We transform our problem into a suitable form to utilize quantum devices. In our case, this amounts to rewriting the given MAX-3SAT instance as a QUBO problem.
We extend the QUBO formulation by Dinneen~\cite{dinneen-qubo} to be able to prioritize satisfying hard over soft constraints.

Consider a clause $$\xi_i = \left(x_{i1}\lor x_{i2} \lor x_{i3}\right),\quad x_{ij}\in\left\{v_1,...,v_m,\Bar{v}_1,...,\Bar{v}_m\right\}.$$
Using the fact that we can represent the negation of binary variables as $\Bar{v}\Leftrightarrow(1-v)$, we can equivalently state the clause $\xi_i$ as a cubic polynomial in the binary variables $x_{ij}$:
\begin{equation}
\label{eq:3sat-pubo}
    \xi_i
    =
    x_{i1} + x_{i2} + x_{i3}
    - x_{i1} x_{i2} - x_{i1}x_{i3} - x_{i2}x_{i3}
    + x_{i1}x_{i2}x_{i3}.
\end{equation}
By introducing an ancillary binary variable $z_i$, we can reduce the degree of the polynomial on the r.h.s. of Eq.~\eqref{eq:3sat-pubo} as
\begin{equation}
\label{eq:cubic-reduction}
    x_{i1}x_{i2}x_{i3}
    = 
    \max_{
    z_i\in\left\{0,1\right\}
    }
    z_i\left(
    x_{i1} + x_{i2} + x_{i3} - 2
    \right).
\end{equation}
As for binary variables $x = x^2$, we can write each clause as a purely quadratic polynomial by combining Eqs.~\eqref{eq:3sat-pubo} and~\eqref{eq:cubic-reduction}.

Let us denote with $\widetilde{\phi}_j(\mathbf{v}, \mathbf{z}_h)$ and $\widetilde{\theta}_k(\mathbf{v}, \mathbf{z}_s)$ quadratic polynomials corresponding to the hard and soft constraints transformed in this manner.
Here $\mathbf{z}_h$ and $\mathbf{z}_s$ are binary vectors of dimensions ${N_\mathrm{h}}$ and ${N_\mathrm{s}}$, with their components being the ancillary variables introduced to reduce the degree of the hard and soft constraints, respectively. 
The vehicle options MAX-3SAT problem can then be formulated as finding the maximum of the following QUBO problem:
\begin{equation}
\label{eq:cmaxsat}
    C_{\mathrm{MAX-SAT}}(\mathbf{v},\mathbf{z}_h, \mathbf{z}_s)
    =
    \lambda\sum_{j=1}^{{N_\mathrm{h}}} \widetilde{\phi}_j(\mathbf{v},\mathbf{z}_h)
    +
    \sum_{k=1}^{{N_\mathrm{s}}} \widetilde{\theta}_k (\mathbf{v},\mathbf{z}_s),
\end{equation}
where $\lambda$ is a hyperparameter.
If we set $\lambda$ to be the number of soft constraints $q$, it is never favorable to violate a hard constraint in order to satisfy a soft constraint.\footnote{If one considers the weighted extension, then we have to set $\lambda\geq\sum_{k=1}^{N_{\mathrm{s}}} w_k$.}
In that case 
\begin{equation}
    \mathbf{v}_{\mathrm{opt}}
    :=
    \argmax_{\mathbf{v}}\left[\max_{\mathbf{z}_h, \mathbf{z}_s}C_{\mathrm{MAX-SAT}}\left(\mathbf{v},\mathbf{z}_h, \mathbf{z}_s\right)\right]
\end{equation}
is guaranteed to be the optimal configuration for the given instance.
Conversely, we can minimize $-C_{\mathrm{MAX-SAT}}$ using (quantum) annealing approaches to obtain $\mathbf{v}_{\mathrm{opt}}$.
Note that this approach uses ${N_\mathrm{f}}+{N_\mathrm{h}}+{N_\mathrm{s}}$ binary variables, and therefore qubits, to encode the vehicle options problem.

While the procedure presented above works for the MAX-3SAT problem, we also include a direct QUBO formulation for MAX-SAT instances with arbitrary (even varying) clause lengths in QUARK.
The formulation relies on mapping the SAT problem to the maximum independent set problem and is an extension of the encoding introduced by Choi~\cite{choi_different_2011}.

\section{QUARK Benchmarking Framework}
\label{sec:quark}

The QUARK framework aims to facilitate the development of application-level benchmarks. The framework simplifies the end-to-end process of designing, implementing, conducting, and communicating application benchmarks. As applications are highly diverse, it is essential to provide a flexible framework that focuses on investigating system performance in terms of application-level quality metrics (e.\,g., the path length for TSP applications), bridging the gap between existing system benchmarks and applications. The framework addresses essential benchmarking requirements, allowing for rapid development and refinement of application benchmarks. It provides reproducibility, verifiability, high usability, and customizability. It ensures that benchmark results can be easily collected and distributed. Furthermore, it is vendor-agnostic, ensuring the neutrality of the system.




\subsection{Architecture}
\label{subsec:general_architecture}

\begin{figure}[t]
    \centering
    \resizebox{0.55\linewidth}{!}{%
    \tikzset{every picture/.style={line width=0.75pt}} 

\begin{tikzpicture}[x=0.75pt,y=0.75pt,yscale=-1,xscale=1]

\draw  [fill={rgb, 255:red, 211; green, 211; blue, 211 }  ,fill opacity=1 ] (148,17) -- (415,17) -- (415,265) -- (148,265) -- cycle ;
\draw  [fill={rgb, 255:red, 255; green, 255; blue, 255 }  ,fill opacity=1 ] (214,22) -- (407,22) -- (407,76) -- (214,76) -- cycle ;
\draw  [fill={rgb, 255:red, 255; green, 255; blue, 255 }  ,fill opacity=1 ] (214,83) -- (407,83) -- (407,137) -- (214,137) -- cycle ;
\draw  [fill={rgb, 255:red, 255; green, 255; blue, 255 }  ,fill opacity=1 ] (214,143) -- (407,143) -- (407,197) -- (214,197) -- cycle ;
\draw  [fill={rgb, 255:red, 255; green, 255; blue, 255 }  ,fill opacity=1 ] (214,204) -- (407,204) -- (407,258) -- (214,258) -- cycle ;
\draw  [fill={rgb, 255:red, 255; green, 255; blue, 255 }  ,fill opacity=1 ] (208.5,22) -- (208.5,258) -- (154.5,258) -- (154.5,22) -- cycle ;

\draw (270,42) node [anchor=north west][inner sep=0.75pt]  [font=\Large] [align=left] {Application};
\draw (280,100) node [anchor=north west][inner sep=0.75pt]  [font=\Large] [align=left] {Mapping};
\draw (287,162) node [anchor=north west][inner sep=0.75pt]  [font=\Large] [align=left] {Solver};
\draw (285,224) node [anchor=north west][inner sep=0.75pt]  [font=\Large] [align=left] {Device};
\draw (172,219.5) node [anchor=north west][inner sep=0.75pt]  [font=\Large,rotate=-270] [align=left] {Benchmark Manager};

\end{tikzpicture}
    }\upp
    \caption{\textbf{Architecture of QUARK:} The framework follows the separation of concerns design principle encapsulating application- and problem-specific aspects, mappings to mathematical formulations, solvers, and hardware.}\upp\upp
    \label{fig:architecture}
\end{figure}
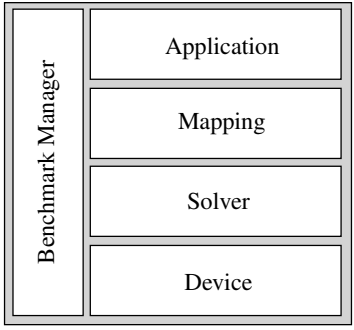


The framework is written in Python and designed to be modular and extensible, facilitating new application and problem types, algorithms, and devices. Figure~\ref{fig:architecture} shows the architecture of the QUARK framework. The framework comprises five components: The \texttt{Benchmark Manager} is responsible for orchestrating the overall execution of the benchmark. The \texttt{Application}, \texttt{Mapping}, \texttt{Solver}, and \texttt{Device} components encapsulate different aspects of a benchmark. 
Each component provides an abstract base class that can be extended for the concrete realizations of a functionality.
The modular approach accommodates changes and extensions to benchmark implementations with minimal effort. 



\emph{Application:}
The application component defines the workload, comprising a dataset of increasing complexity, a validation, and an evaluation function. Generally, the framework can integrate any dataset, e.\,g., real-world, synthetic and established benchmark data (e.\,g., TSPLib95~\cite{RePEc:inm:orijoc:v:3:y:1991:i:4:p:376-384}). 
The application module can be configured using a shared, framework-wide configuration management system, e.\,g., different problem sizes can be generated depending on the configuration, accommodating the limitations of current quantum hardware and simulation devices. 
The validation function checks whether the provided solution is valid. For example, the function determines whether a valid path comprising a visit of all seams was generated for the robot path problem. The validation function assumes that the result can be validated using a classical system, which is the case for most problems. 
The task of the evaluation function is to compute and return a metric that aids the quantitative comparison of the discovered solution. The benchmark developer can utilize particular quality scores for this purpose. 


\emph{Mapping:}
The task of the mapping module is to translate the application's data and problem specification into a mathematical formulation suitable for a solver. For example, quantum-based solvers for combinatorial optimization problems usually require the problem to be specified in a QUBO or Ising formulation~\cite{DBLP:journals/corr/abs-1811-11538}. 
The mapping is highly application-specific, requiring domain-specific knowledge. To implement the mapping, developers can utilize higher-level abstractions, e.\,g., PyQubo~\cite{zaman2021pyqubo}, or re-use available formulations in libraries, such as Ocean~\cite{ocean} and Qiskit Optimization~\cite{qiskit-optimize}.



\emph{Solver:}
The solver is responsible for finding feasible and high-quality solutions to the formulated problem, i.\,e., of the defined objective function. Various algorithms for solving QUBO problems exist, e.\,g., quantum annealing as provided by D-Wave machines, QAOA~\cite{farhi2014quantum} and VQE~\cite{vqe_2014} for NISQ devices, and Grover Adaptive Search for fault-tolerant hardware~\cite{bulger_implementing_2003}.  
Quantum SDKs like Qiskit~\cite{qiskit} and PennyLane~\cite{bergholm2020pennylane} provide circuit templates or higher abstractions for solving QUBO and Ising problem formulations. 



\emph{Device:}
Several quantum devices (e.\,g., IonQ, Rigetti, IBM, Google), simulators (e.\,g.,  Amazon Braket's SV1, Qiskit's QASM, and PennyLane's lightning simulator), and services (e.\,g., Amazon Braket and Azure Quantum) exist. Each environment has its characteristics and API. Adapting applications and benchmarks to this heterogeneous landscape is challenging, requiring the manual customization of API (e.\,g., for job submission) and translation between data formats (e.\,g., different  QUBO/Ising matrix representations). 

The device class abstracts details of the physical device, such as submitting a task to the quantum system.
QUARK currently supports different simulators, e.\,g.,  Amazon Braket, Qulacs, and Qiskit, and quantum hardware, i.\,e., annealing, gate-based superconducting and ion-trap based quantum computers via Amazon's Braket service. It can easily be extended to additional simulators and quantum hardware systems.


\emph{Benchmark Manager:}
The benchmark manager is the main component of QUARK orchestrating the overall benchmarking process. The benchmarking process is highly customizable, i.\,e., every module is configurable using a central configuration file. Custom parameter settings can be added for all components, allowing a straightforward evaluation of different parameters. This configuration system ensures that benchmarks and parameters can easily be standardized.
Based on the configuration, the benchmark manager will create an experimental plan considering all combinations of configurations, e.\,g., different problem sizes, solver, and hardware combinations. It will then instantiate the respective framework components representing the application, the mapping to the algorithmic formulation, solver, and device. 

After executing the benchmarks, QUARK collects the generated data and executes the validation and evaluation functions. Data is processed according to the tidy specification~\cite{Wickham:2014:TidyData} and stored with its metadata, such as the used configuration, to ensure reproducibility. Further, the framework creates various analysis plots automatically. The well-defined data model can also accommodate manual data analytics, e.\,g., for profiling.

\begin{figure}[t]
    \centering
    \resizebox{0.9\linewidth}{!}{%
    \tikzset{every picture/.style={line width=0.75pt}} 

\begin{tikzpicture}[x=0.75pt,y=0.75pt,yscale=-1,xscale=1]

\draw   (39.8,182.6) -- (149.2,182.6) -- (149.2,213) -- (39.8,213) -- cycle ;
\draw   (180.6,182) -- (290,182) -- (290,212.4) -- (180.6,212.4) -- cycle ;
\draw   (320.2,263.4) -- (429.6,263.4) -- (429.6,293.8) -- (320.2,293.8) -- cycle ;
\draw   (320.2,105.2) -- (429.6,105.2) -- (429.6,135.6) -- (320.2,135.6) -- cycle ;
\draw   (461,203) -- (570.4,203) -- (570.4,233.4) -- (461,233.4) -- cycle ;
\draw   (460.6,263.8) -- (570,263.8) -- (570,294.2) -- (460.6,294.2) -- cycle ;
\draw   (460.6,136) -- (570,136) -- (570,166.4) -- (460.6,166.4) -- cycle ;
\draw   (460.6,74) -- (570,74) -- (570,104.4) -- (460.6,104.4) -- cycle ;
\draw  [dash pattern={on 0.84pt off 2.51pt}]  (165.2,46) -- (166.5,362) ;
\draw  [dash pattern={on 0.84pt off 2.51pt}]  (304.8,45.6) -- (304.5,361) ;
\draw    (150.8,197.6) -- (179.2,197.6) ;
\draw    (292.5,198) -- (315.5,120) ;
\draw    (292.5,198) -- (317.5,278) ;
\draw    (458.4,150.8) -- (431.2,120.8) ;
\draw    (458,88.4) -- (431.2,120.8) ;
\draw    (458.5,218) -- (432.4,277.4) ;
\draw    (457.5,277) -- (432.4,277.4) ;
\draw  [dash pattern={on 0.84pt off 2.51pt}]  (445.2,45.6) -- (445.5,360) ;
\draw   (461.6,323.8) -- (571,323.8) -- (571,354.2) -- (461.6,354.2) -- cycle ;
\draw    (459.5,340) -- (432.4,277.4) ;

\draw (79.2,191.4) node [anchor=north west][inner sep=0.75pt]   [align=left] {PVC};
\draw (214.4,190.4) node [anchor=north west][inner sep=0.75pt]   [align=left] {QUBO};
\draw (353.4,271) node [anchor=north west][inner sep=0.75pt]   [align=left] {QAOA};
\draw (342.6,112) node [anchor=north west][inner sep=0.75pt]   [align=left] {Annealing};
\draw (500.8,212.2) node [anchor=north west][inner sep=0.75pt]   [align=left] {SV1};
\draw (495.2,271) node [anchor=north west][inner sep=0.75pt]   [align=left] {Rigetti};
\draw (483.8,143.2) node [anchor=north west][inner sep=0.75pt]   [align=left] {Simulator};
\draw (489.6,81.6) node [anchor=north west][inner sep=0.75pt]   [align=left] {D-Wave};
\draw (354.4,35.6) node [anchor=north west][inner sep=0.75pt]   [align=left] {Solver};
\draw (206.4,35.6) node [anchor=north west][inner sep=0.75pt]   [align=left] {Mapping};
\draw (58,35.6) node [anchor=north west][inner sep=0.75pt]   [align=left] {Application};
\draw (492.4,35.6) node [anchor=north west][inner sep=0.75pt]   [align=left] {Device};
\draw (496.2,332) node [anchor=north west][inner sep=0.75pt]   [align=left] {Qulacs};

\end{tikzpicture}}\upp
    \caption{Example of how an application can be combined with different mappings, solvers and devices.\upp\upp}
    \label{fig:appMapSolvDev}
\end{figure}
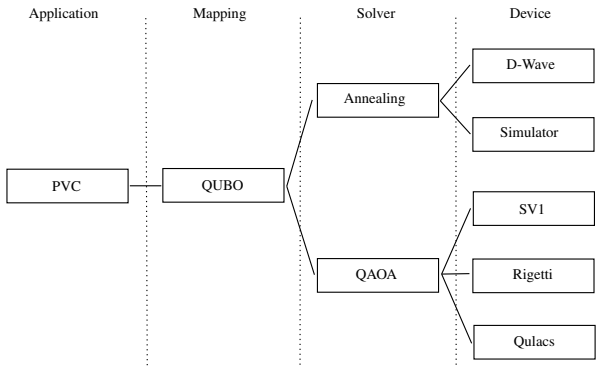

Figure~\ref{fig:appMapSolvDev} illustrates an example of concrete instances of the abstract components. For example, the robot path planning application generates a synthetic application graph mimicking real-world data and stores it as a NetworkX graph object. The current implementation provides different mapping options, e.\,g., a custom or a predefined (from e.\,g. Qiskit) QUBO mapping.
The QUBO formulation is then used to solve the problem using quantum annealing, QAOA or classical methods like simulated annealing. The device abstraction provides the means to execute application tasks.

\subsection{Key Metrics}
\label{subsec:def_kpis}
Defining relevant metrics is one of the key challenges when creating benchmarks. QUARK supports a set of well-defined metrics that in particular attempt to balance the trade-offs between the time-to-solution $TTS$, the validity $V$, and the quality $Q$ of a solution.
$V$ indicates whether a solution found by the solver is valid (e.g. if it conforms to all constraints). Both $V$ and $Q$ are application-specific and can be customized.

$TTS$ is defined as the end-to-end time required to obtain a solution. It is decomposed into several components:
\begin{equation}
\label{eq:metrics}
\begin{aligned}
TTS ={} & T_{\mathrm{mapping}} + T_{\mathrm{solver}} + T_{\mathrm{reverseMap}} \\
      & + T_{\mathrm{processSolution}} + T_{\mathrm{validation}} + T_{\mathrm{evaluation}}.
\end{aligned}
\end{equation}
Here, $T_{\mathrm{solver}}$ denotes the runtime for the solver with a given configuration. $T_{\mathrm{mapping}}$ gives the time required to map an application formulation to a representation required by the solver, e.\,g., the time required to convert a graph into a QUBO instance. $T_{\mathrm{reverseMap}}$ and $T_{\mathrm{processSolution}}$ are the execution times of two intermediate steps, needed to convert the solution to a representation that can be used for validation and evaluation. We store the time elapsed during validation and evaluation as $T_{\mathrm{validation}}$ and $T_{\mathrm{evaluation}}$, respectively.

\section{Performance Characterization}
\label{sec:results}

\note{Reviewers: The experimental results could also include other quantum computers than D-Wave
3. The experiment evaluation should include a breakdown between the time executed on the quantum computer and a classical one.}


We demonstrate the capabilities of QUARK by applying it to the applications presented in Section~\ref{sec:applications}.  We present some initial results for these applications.  The intention of these results is not to highlight the best approach to solve a given problem but to showcase the flexibility of QUARK, and the value of providing real-world applications. 


\subsection{Experimental Setup}

All non-quantum operations were executed on an \emph{NVIDIA DGX A100} device (Dual AMD Rome 7742, 2\,TB memory, 8x NVIDIA A100 40\,GB). We only use the GPU for selected experiments.
Every experiment configuration is repeated at least five times to compute a variability measure. Problem sizes are chosen according to the current capabilities of quantum devices.  While we have conducted some micro-experiments to identify suitable configurations of hyperparameters, we focused on understanding out-of-the-box performance rather than deeply profiling a single configuration.



We investigate different classical solvers and D-Wave quantum annealers for all applications. For TSP, we also perform classical simulations of QAOA. We assess quantum annealing on the two D-Wave machines available on Amazon Braket: (\emph{D-Wave Advantage 4.1} with 5760 Qubits and \emph{2000Q 6} with 2048 Qubits).
It is insightful to compare quantum annealing to its classical counterpart, simulated annealing -- we use the implementation from the Neal library~\cite{simannealsampler}.
For all annealing methods, we used 500 reads.
Although a QUBO formulation is typically not the most efficient mathematical representation for simulated annealing, this approach aids a direct comparison between quantum and simulated annealing. 



     


We investigate the solution validity $V$, quality $Q$, and time-to-solution $TTS$. In all experiments, $TTS$ is mainly determined by $T_{\mathrm{solver}}$. The other components of $TTS$ do not significantly change for different problem sizes. For example, the annealing of the robot path problem $T_{\mathrm{solver}}$ accounts for more than $99\%$ of the overall $TTS$. As for quantum annealing, $T_{\mathrm{solver}}$ also includes the embedding time, which significantly impacts $TTS$ for larger problem instances (e.\,g., for both TSP and PVC, the embedding time accounts for around $80\%$ of the overall $TTS$ for the largest problem size). The error bars (where visible) display the minima and maxima across different solver runs. 


\subsection{Robot Path Planning (PVC Sealing)}

\begin{figure}
    \includegraphics[width=\linewidth]{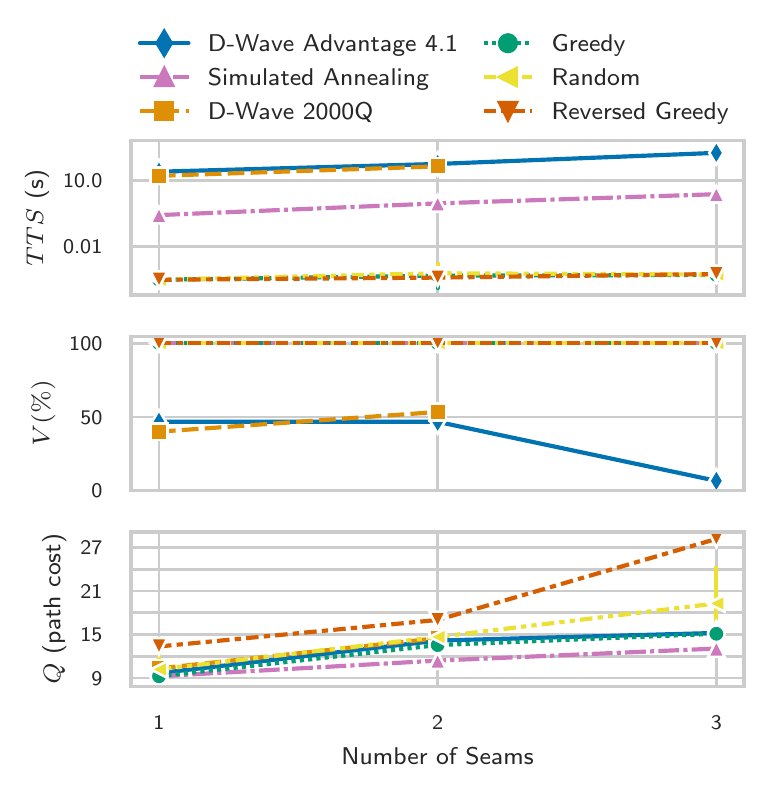}\upp\upp
    

    \caption{\textbf{Robot Path Optimization -- Annealing Results} for different number of seams.   \label{fig:pvc_plot}  \label{fig:pvc_lineplot} Simulated annealing achieves the best solution quality (bottom panel; lower is better) on average, while both D-Wave devices struggle to find valid solutions (middle panel).\upp\upp
    }
\end{figure}

Figure~\ref{fig:pvc_plot} summarizes the results for $TTS$, the path length $Q$, and for the ratio of valid solutions $V$. A path is valid if it starts from the home position and visits all seams.
In addition to simulated annealing, we implemented three other classical algorithms as baselines: greedy, reversed greedy, and random.
The greedy and reversed greedy algorithms make the best and worst possible local move at each step, respectively. 
The random solver makes a random choice at every time step to decide which node to visit next.


While quantum annealing outperforms the reverse greedy and random algorithms, it performs worse than simulated annealing -- particularly striking is the difference between the ratios of valid solutions.
As both simulated and quantum annealing use the same problem formulation, this suggests that the capabilities of available quantum devices, rather than the problem formulation and encoding, are the limiting factors.

Another limitation of current D-Wave devices is that embedding larger problem sizes is impossible after a few seams.  On  \emph{D-Wave 2000Q} we can only solve two seams, while on the larger \emph{Advantage 4.1}, problems up to three seams can be solved. It is possible, however, that a QUBO formulation tailored to the particular architecture of D-Wave devices, would perform significantly better, both in terms of the solution quality and the time-to-solution.




\subsubsection*{Traveling Salesperson}

The TSP problem represents a simplification of the PVC sealing problem. In the following, we use TSP to establish baseline for the PVC experiments. By integrating the TSPLib95~\cite{RePEc:inm:orijoc:v:3:y:1991:i:4:p:376-384} into QUARK, we can benchmark quantum TSP solutions against state-of-the-art solutions.

\begin{figure}[t]
    \includegraphics[width=\linewidth]{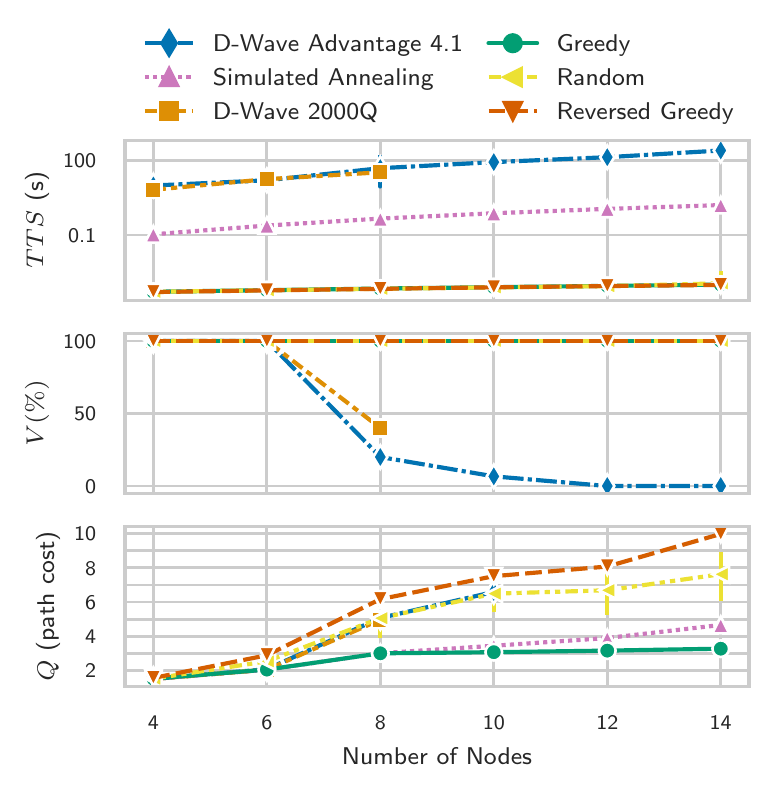}\upp\upp
    
    \caption{\textbf{TSP -- Annealing Results} for different numbers of nodes.  \label{fig:tsp_plot} \label{fig:tsp_lineplot}
    While on average, the greedy solver achieves a better solution quality (bottom panel; lower is better), we find, up to eight nodes, at least one annealing run with a better solution. Starting at 8 nodes, we observe a drop in the rate of valid solutions (middle panel) for both quantum annealers.\upp\upp}
\end{figure}

Fig.~\ref{fig:tsp_plot} illustrates the performance obtained using the dsj1000 TSPLib95 dataset, which we reproducibly simplified by removing nodes until reaching the desired problem size. The QUBO formulation for this problem is constructed from the graph using the Ocean library~\cite{ocean}, and requires $N_{\mathrm{nodes}}^2$ qubits. We compare quantum and simulated annealing to different classical algorithms: NetworkX's greedy algorithm~\cite{networkX}, the described reversed greedy and random algorithms. 



On average, the greedy solver returns the shortest paths. However, for up to eight nodes, we always find at least one annealing run with a better solution (e.\,g., all three annealing options for six nodes). Thus, it is important to not only consider the average performance.

In summary, simulated annealing exhibits a better performance than quantum annealing. 
However, while for PVC sealing, simulated annealing outperforms the greedy algorithm, the opposite is true for the TSP. The reason is that the greedy algorithm for PVC sealing never changes its tool and config setting during a tour, as it is never locally optimal to do so.

As for PVC sealing, above a certain problem size, finding an embedding for quantum annealers is impossible.
On the \emph{D-Wave 2000Q} we can solve problems with at most eight nodes, while on the larger \emph{D-Wave Advantage 4.1} instances with up to 14 nodes are feasible. Furthermore, starting at eight nodes, we observe a drop in the ratio of valid solutions for both quantum annealers.
Finally, for more than ten nodes, no valid solutions could be found with \emph{D-Wave Advantage 4.1}.



\subsubsection*{Variational Algorithms}


QAOA~\cite{farhi2014quantum} is a variational quantum algorithm suitable for NISQ devices. 
It provides approximate solutions to combinatorial optimization problems. 
To assess the performance of QAOA, we implemented it with different libraries (e.\,g., Amazon Braket~\cite{braket}, PennyLane~\cite{bergholm2020pennylane}, and Qiskit~\cite{qiskit}).\alnote{maybe we could speak to some of the challenges in implementing / training QAOA} Here, we present the results of a PennyLane-based QAOA implementation evaluated using QUARK on different CPU and GPU-based simulators.
For example, PennyLane's \emph{lightning.qubit} is a CPU-based simulator with built-in parallelization; \emph{lightning.gpu} utilizes the cuStateVec library to offload computations to the GPU (currently, only one GPU is supported).
We evaluated the performance on TSP instances up to five nodes. For all experiments, $60$ iterations were performed, using the adjoint differentiation method~\cite{jones}, and the momentum optimizer (stepsize: $0.001$, momentum: $0.9$).

\begin{figure}[t]
    \centering
    \includegraphics[width=\linewidth]{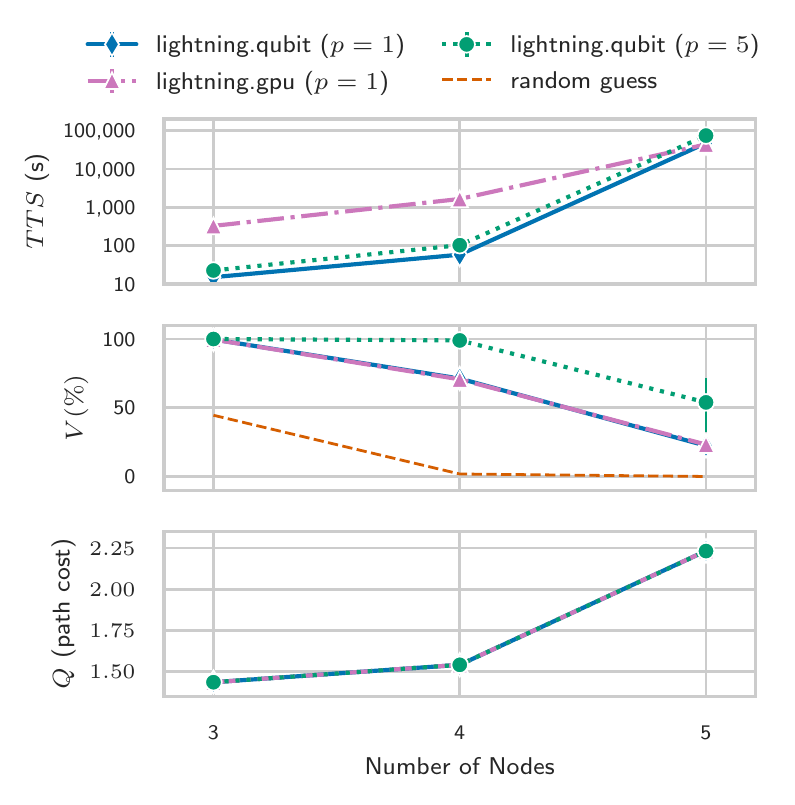}\upp\upp
    \caption{\textbf{TSP -- QAOA Results} for different numbers of nodes and CPU and GPU simulators. The trend of the $TTS$ (upper panel) indicates that GPU simulators could outperform CPU simulators at larger system sizes, while CPU-based simulations are more efficient at smaller system sizes. Validity (middle panel) improves slightly with a higher number of QAOA layers $p$, while the quality curves (bottom panel; lower is better) collapse, as the algorithm cannot differentiate between the different valid solutions.
    \jrfnote{$p$ is the number of layers comment.}\upp\upp}
    \label{fig:qaoa_exp}
\end{figure}

\begin{figure}[t]
    \includegraphics[width=0.98\linewidth]{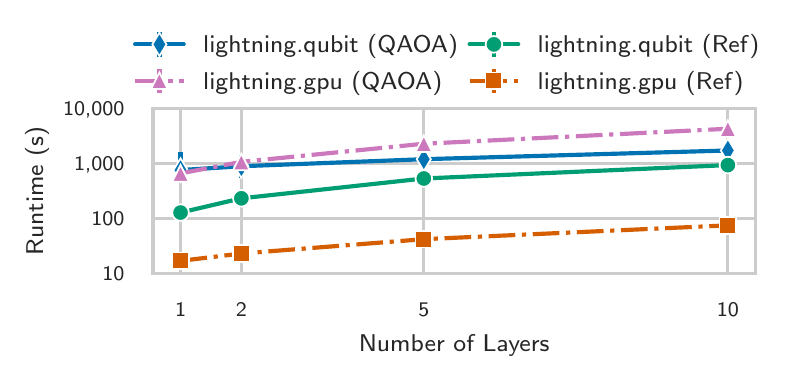}\upp\upp
    \caption{\textbf{QAOA Results -- Runtime for a single circuit execution} of QAOA and PennyLane's \emph{StronglyEntanglingLayers} benchmark~\cite{pennylaneGPU} (Ref). Both scenarios use 25 qubits.
    While the execution of Pennylane's reference circuit is significantly faster on a GPU than on a CPU, the converse holds for a QAOA circuit.\upp\upp}
    \label{fig:qaoa_tts}
\end{figure}

The top panel in Fig.~\ref{fig:qaoa_exp} shows that the CPU-based simulator has a better $TTS$ than the GPU simulator, particularly for small instances. The runtime on CPU and GPUs are comparable for five node instances (corresponding to $25$ qubits). To investigate the lack of performance gains using the GPU simulator, we conducted a micro experiment evaluating a single circuit execution of QAOA, averaged over ten runs (Fig.~\ref{fig:qaoa_tts}). We compare our QAOA circuit to PennyLane's \emph{StronglyEntanglingLayers} benchmark~\cite{pennylaneGPU}. The experiment confirmed that for the problem sizes considered our QAOA circuit does not benefit from the GPU acceleration as the overheads (e.g., data transfer, synchronization) negate a potential computational advantage. Further, the CPU simulator heavily utilizes the parallelism available on the $64$ CPU cores machine. We expect that the GPU support will improve significantly in the future (see Pennylane's lightning roadmap~\cite{pennylaneGPU}).


We define the validity $V$ as the probability of obtaining a valid bitstring (i.\,e., one that corresponds to an actual tour) among 50 measurements of the output state,\alnote{are we doing here measurements or are we utilizing the resulting probability distribution? What is the performance impact of both approaches?}\jrfnote{We are utilizing the probability distribution. It would be more or less useless to sample from the circuit if we're doing a state-vector simulation.}\alnote{so why do we do measurements then?} averaged over five runs of the algorithm. \alnote{can we speak to the challenges in particular the hyperparameter optimization required to obtain these results?} As shown in the middle panel of Fig.~\ref{fig:qaoa_exp}, QAOA is able to identify valid solutions. However, $V$ decreases with increasing problem size, i.\,e., the number of nodes. The data also indicates that this issue could be remedied by deeper QAOA circuits; in our experiments, $V$ increased slightly for five layers.\alnote{can we reference here examples that indicate the challenges with scaling/converging QAOA for problems other than Max-Cut?}

The bottom panel of Fig.~\ref{fig:qaoa_exp} displays the quality metric $Q$, defined as the\alnote{is this the expectation value or the mean of all valid paths?} expectation value of the path cost. $Q$ collapses to the \alnote{we should consistently refer to mean, average or expectation value} mean tour length for both one and five layers.  
This could indicate that the Lagrange parameter value (here set to twice the average tour length) is too large. Our extensive hyperparameter search indicated a trade-off between $V$ and $Q$ and that smaller Lagrange parameter values lead to a decrease in $V$.

Beyond tuning hyperparameters, potential improvements to the performance of QAOA on TSP could come from improving the TSP formulation. The current formulation requires $N_{\mathrm{nodes}}^2$ qubits and in turn leads to numerous spurious states which do not correspond to a valid tour.
As getting the algorithm to converge to valid solutions was rather challenging due to the constrained nature of the problem, we will investigate other initialization schemes (e.\,g., warm-starting~\cite{Egger_2021} or Dicke state initialization~\cite{qpack}) and constrained mixers~\cite{alteroper} to decrease the reliance on the Lagrange parameter for constraint enforcement.

\subsection{Vehicle Options}


We evaluate the vehicle options inspired instances of MAX-3SAT using randomly generated MAX-3SAT instances for a range of total feature (variable) numbers $N_\mathrm{f}$ up to $110$, which is the largest problem instance we can encode on a quantum annealer. 
For each $N_\mathrm{f}$, we generate ten different MAX-3SAT instances with $N_\mathrm{h}=2N_\mathrm{f}$ hard constraints and $N_\mathrm{s}=\ceil{4.2N_\mathrm{f}}$ soft constraints.
Additionally, we ensure that no variable appears more than once within each clause.
 
We utilize the QUBO formulation presented in Section~\ref{sec:vehicle-options} (using $\lambda={N_\mathrm{s}}$) to solve these problems using two different quantum annealing devices and a classical simulated annealing algorithm.
With the given problem specifics, the number of qubits needed to encode the generated instances scales as linearly as $\ceil{7.2N_\mathrm{f}}$.
We benchmarked the annealing-based approaches against the designated classical MAX-SAT solver RC2~\cite{sat-solver-rc2}.
We perform three solver runs for each problem instance, resulting in $30$ runs per solver for each $N_{\mathrm{f}}$.

\begin{figure}[t]
    \includegraphics{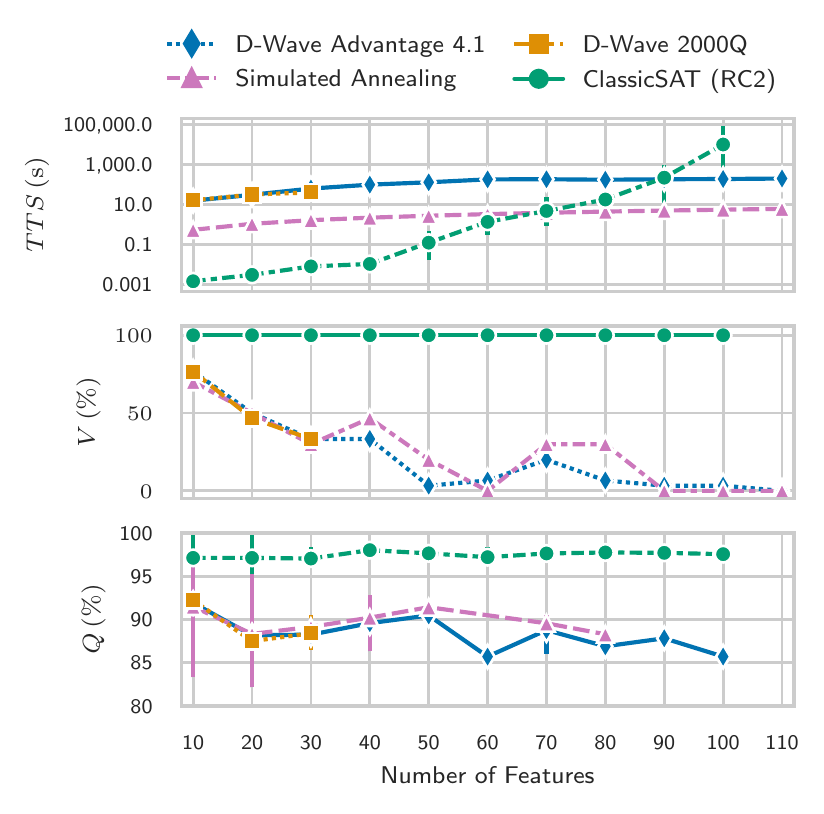}\upp\upp

    \caption{\textbf{Vehicle Options -- Annealing Results.}
    With increasing instance complexity, the classical solver's $TTS$ (upper panel) increases exponentially, while the scaling appears to be subexponential for annealing approaches.
    However, the classical solver outperforms annealing methods regarding the ratio of valid solutions (middle panel) and the quality of returned solutions (bottom panel).
    For larger problem instances, annealing algorithms only sporadically return valid solutions, showing a noticeable decline in performance.\upp\upp
    \label{fig:sat-plot}}

\end{figure}

In Fig.~\ref{fig:sat-plot} we display the $TTS$, the ratio and average quality of valid solutions returned by each solver. The quality metric is defined as the ratio of satisfied soft constraints and is only displayed for valid solutions, i.\,e., solutions that satisfy all hard constraints.
These results reveal several features of the solvers we analyzed.
Firstly, we observed that annealing-based approaches do not consistently return valid solutions -- this would suggest that increasing the $\lambda$ parameter (see Eq.~\eqref{eq:cmaxsat}) is required. 
However, the ratio of satisfied soft constraints roughly coincides with that expected from random assignments, which is $87.5\%$.\footnote{Notice there is a single invalid configuration for each clause. Hence, a random assignment has a probability of $1-1/2^3=87.5\%$ to satisfy each clause (of length 3).}
This suggests that the annealing methods completely disregard soft constraints -- increasing $\lambda$ would only exacerbate this problem. 
This issue often arises in constrained QUBO formulations as QUBOs are inherently unconstrained.
Hence, one has to carefully balance enforcing constraints and optimizing the objective~\cite{Lucas_2014}.

Secondly, there is a 
big difference in solution quality between the RC2 classical solver and the annealing-based approaches. However, the $TTS$ of RC2 increases roughly exponentially, especially for larger problem sizes ($N_\mathrm{f}\geq 40$).
While more efficient approximate classical algorithms exist~\cite{Joshi2018}, this gives hope that quantum annealing could become a viable alternative with improved encoding and devices.
Such improvements could come from tuning hyperparameters (e.\,g., $\lambda$) of the QUBO mappings presented within our framework or from finding more efficient encodings that potentially better suit the topology of current annealing devices~\cite{Bian2020, Chancellor2016}.

While one (on average) expects a monotonic decrease in performance of annealing algorithms with increasing problem sizes~\cite{znidaric2006, Azinovic2017}, this is not strictly the case in our study (see middle panel of Figure~\ref{fig:sat-plot}).
This behavior can be explained by the fact that we generate a limited number of instances at each $N_\mathrm{f}$.
These instances can, in principle, be of varying complexity, which in turn leads to the varying performance of the solvers -- the trend towards worsening efficacy as the problem sizes increase is, however, evident.
Varying instance complexity manifests itself in fluctuating solution validity for annealing-based approaches and in the variation of $TTS$ for the classical solver 
(note error bars in Fig~\ref{fig:sat-plot}'s top panel).


Finally, we can observe that the quantum and simulated annealing approaches yield comparable results.
Moreover, it is interesting to note that quantum annealing provided valid solutions for some problem sizes where no valid solution was found with simulated annealing.
The fact that, at least for this use case, quantum annealing seems to have started catching up with its classical counterpart portends optimism as quantum annealing devices are improved.

\section{Conclusion and Future Work}
\label{sec:conclusion}

Benchmarks are instrumental for measuring progress, encouraging new and innovative solutions, accelerating adoption, establishing best practices, and predicting the viability of algorithms and hardware solutions. 
In this paper, we make a case for application-centric benchmarks to connect progress in the QC hardware realm to real-world application performance. 
For this purpose, we propose a ``pen and paper'' benchmark approach to address the uncertainty concerning practical quantum advantages.
QUARK automates and standardizes critical parts of a benchmarking system, ensuring reproducibility and verifiability.
The modular architecture enables benchmark developers to investigate and automate large-scale benchmark scenarios across diverse infrastructures.
We envision that a wide variety of community-driven benchmarks will guide the progress toward a practical advantage for industrial quantum applications. QUARK-based benchmarks will help quantum application developers to identify performance bottlenecks, compare different algorithms, hardware and software configurations, and estimate resource requirements.


We demonstrate the benchmark development lifecycle from specification, implementation to execution using QUARK using two significant and representative industrial applications: \emph{robot path} and \emph{vehicle option} optimization.  Our results provide valuable insights into the current state of quantum computing. Unsurprisingly, classical solvers outperform quantum algorithms in that they more reliably return valid and higher quality solutions.
However, the roughly exponential scaling of the $TTS$ for the classical solver in the vehicle options problem emphasizes the opportunity for a potential quantum computing advantage.
While our results show limitations of current quantum approaches, we believe QUARK will 
be valuable for advancing application benchmarks.









\emph{Future Work:} We will evolve QUARK by adding new problem classes (e.\,g., machine learning and chemistry) and frameworks (e.\,g., Amazon Braket Hybrid Jobs). Particularly, we will add the functionality of comprehensively analyzing hybrid algorithms, facilitating the in-depth characterization of all classical and quantum components. Further, we will enrich the data and metrics, e.g., by supporting lower-level metrics like gate fidelities to understand the system's behavior better.

We will evolve the presented reference implementations into standardized benchmarks. Standardizing all aspects of benchmarks is crucial to advance the uptake, utility, and impact. In addition to technical aspects, engaging interested parties in a community-driven process of the technology industry, application users, and academia is crucial.



\small
\section*{Acknowledgment}
\footnotesize
\noindent
We thank S. Benesch, Y. van Dijk, M. Erdmann, C. Mendl, L. Müller and C. Riofrío for valuable feedback. Additionally, we thank AWS, specifically K. Brubaker, H. Katzgraber, H. Montagu, M. Resende, and M. Schuetz for the TSP QUBO formulation. PR and JK are partly funded by the German Ministry for Education and Research (BMBF) (Project: QAI2-Q-KIS/\#13N15587).
\bibliographystyle{IEEEtran}
\bibliography{main,references}

\begin{thebibliography}{10}
\providecommand{\url}[1]{#1}
\csname url@samestyle\endcsname
\providecommand{\newblock}{\relax}
\providecommand{\bibinfo}[2]{#2}
\providecommand{\BIBentrySTDinterwordspacing}{\spaceskip=0pt\relax}
\providecommand{\BIBentryALTinterwordstretchfactor}{4}
\providecommand{\BIBentryALTinterwordspacing}{\spaceskip=\fontdimen2\font plus
\BIBentryALTinterwordstretchfactor\fontdimen3\font minus
  \fontdimen4\font\relax}
\providecommand{\BIBforeignlanguage}[2]{{%
\expandafter\ifx\csname l@#1\endcsname\relax
\typeout{** WARNING: IEEEtran.bst: No hyphenation pattern has been}%
\typeout{** loaded for the language `#1'. Using the pattern for}%
\typeout{** the default language instead.}%
\else
\language=\csname l@#1\endcsname
\fi
#2}}
\providecommand{\BIBdecl}{\relax}
\BIBdecl

\bibitem{mck_2021}
M.~Biondi, A.~Heid \emph{et~al.}, ``Quantum computing: An emerging ecosystem
  and industry use cases,'' McKinsey \& Company,
  \url{https://www.mckinsey.com/business-functions/mckinsey-digital/our-insights/quantum-computing-use-cases-are-getting-real-what-you-need-to-know},
  2021.

\bibitem{qutac_epj}
\BIBentryALTinterwordspacing
A.~Bayerstadler, G.~Becquin \emph{et~al.}, ``Industry quantum computing
  applications,'' \emph{EPJ Quantum Technology}, vol.~8, no.~1, p.~25, Nov.
  2021. [Online]. Available:
  \url{https://doi.org/10.1140/epjqt/s40507-021-00114-x}
\BIBentrySTDinterwordspacing

\bibitem{Arute2019}
\BIBentryALTinterwordspacing
F.~Arute, K.~Arya \emph{et~al.}, ``Quantum supremacy using a programmable
  superconducting processor,'' \emph{Nature}, vol. 574, no. 7779, pp. 505--510,
  Oct. 2019. [Online]. Available:
  \url{https://doi.org/10.1038/s41586-019-1666-5}
\BIBentrySTDinterwordspacing

\bibitem{xanadu}
\BIBentryALTinterwordspacing
L.~S. Madsen, F.~Laudenbach \emph{et~al.}, ``Quantum computational advantage
  with a programmable photonic processor,'' \emph{Nature}, vol. 606, no. 7912,
  pp. 75--81, 2022. [Online]. Available:
  \url{https://doi.org/10.1038/s41586-022-04725-x}
\BIBentrySTDinterwordspacing

\bibitem{1201189}
S.~Sim, S.~Easterbrook, and R.~Holt, ``Using benchmarking to advance research:
  a challenge to software engineering,'' in \emph{25th International Conference
  on Software Engineering, 2003. Proceedings.}, 2003, pp. 74--83.

\bibitem{bcg2022}
M.~Langione, J.-F. Bobier \emph{et~al.}, ``The race to quantum advantage
  depends on benchmarking,''
  \url{https://www.bcg.com/publications/2022/value-of-quantum-computing-benchmarks},
  2022.

\bibitem{PhysRevA.100.032328}
\BIBentryALTinterwordspacing
A.~W. Cross, L.~S. Bishop \emph{et~al.}, ``Validating quantum computers using
  randomized model circuits,'' \emph{Phys. Rev. A}, vol. 100, p. 032328, Sep
  2019. [Online]. Available:
  \url{https://link.aps.org/doi/10.1103/PhysRevA.100.032328}
\BIBentrySTDinterwordspacing

\bibitem{wack2021quality}
A.~Wack, H.~Paik \emph{et~al.}, ``Quality, speed, and scale: three key
  attributes to measure the performance of near-term quantum computers,'' 2021.

\bibitem{lubinski2021applicationoriented}
T.~Lubinski, S.~Johri \emph{et~al.}, ``Application-oriented performance
  benchmarks for quantum computing,'' 2021.

\bibitem{martiel2021benchmarking}
S.~Martiel, T.~Ayral, and C.~Allouche, ``Benchmarking quantum co-processors in
  an application-centric, hardware-agnostic and scalable way,'' 2021.

\bibitem{Bailey2011}
\BIBentryALTinterwordspacing
D.~Bailey, E.~Barszcz \emph{et~al.}, ``The nas parallel benchmarks,'' \emph{The
  International Journal of Supercomputing Applications}, vol.~5, no.~3, pp.
  63--73, 1991. [Online]. Available:
  \url{https://doi.org/10.1177/109434209100500306}
\BIBentrySTDinterwordspacing

\bibitem{quarkGithub}
\BIBentryALTinterwordspacing
(2022) Quark: A framework for quantum computing application benchmarking.
  [Online]. Available: \url{https://github.com/BMW-Group-Quantum/QUARK}
\BIBentrySTDinterwordspacing

\bibitem{ibm}
\BIBentryALTinterwordspacing
{{IBM}}. (2022) Ibm quantum. [Online]. Available:
  \url{https://quantum-computing.ibm.com/}
\BIBentrySTDinterwordspacing

\bibitem{google}
\BIBentryALTinterwordspacing
Google. (2022) Quantum computer datasheet. [Online]. Available:
  \url{https://quantumai.google/hardware/datasheet/weber.pdf}
\BIBentrySTDinterwordspacing

\bibitem{rigetti}
\BIBentryALTinterwordspacing
Rigetti. (2022) Rigetti website. [Online]. Available:
  \url{https://www.rigetti.com/get-quantum}
\BIBentrySTDinterwordspacing

\bibitem{ionq}
\BIBentryALTinterwordspacing
IonQ. (2022) Ionq website. [Online]. Available: \url{https://ionq.com/}
\BIBentrySTDinterwordspacing

\bibitem{honeywell}
\BIBentryALTinterwordspacing
Honeywell. (2022) Honeywell. [Online]. Available:
  \url{https://www.honeywell.com/us/en/company/quantum}
\BIBentrySTDinterwordspacing

\bibitem{dwave}
\BIBentryALTinterwordspacing
{{D-Wave}}. (2022) D-wave: Leap. [Online]. Available:
  \url{https://cloud.dwavesys.com/}
\BIBentrySTDinterwordspacing

\bibitem{lukin-neutral}
\BIBentryALTinterwordspacing
S.~Ebadi, T.~T. Wang \emph{et~al.}, ``Quantum phases of matter on a 256-atom
  programmable quantum simulator,'' \emph{Nature}, vol. 595, no. 7866, pp.
  227--232, Jul. 2021. [Online]. Available:
  \url{https://doi.org/10.1038/s41586-021-03582-4}
\BIBentrySTDinterwordspacing

\bibitem{topological-qc}
\BIBentryALTinterwordspacing
V.~Lahtinen and J.~K. Pachos, ``{A Short Introduction to Topological Quantum
  Computation},'' \emph{SciPost Phys.}, vol.~3, p. 021, 2017. [Online].
  Available: \url{https://scipost.org/10.21468/SciPostPhys.3.3.021}
\BIBentrySTDinterwordspacing

\bibitem{qc_sim_list}
\BIBentryALTinterwordspacing
(2022) List qc simulators. [Online]. Available:
  \url{https://www.quantiki.org/wiki/list-qc-simulators}
\BIBentrySTDinterwordspacing

\bibitem{ferrari1978computer}
\BIBentryALTinterwordspacing
D.~Ferrari, \emph{Computer Systems Performance Evaluation}.\hskip 1em plus
  0.5em minus 0.4em\relax Prentice-Hall, 1978. [Online]. Available:
  \url{https://books.google.de/books?id=geBQAAAAMAAJ}
\BIBentrySTDinterwordspacing

\bibitem{Jain:1991:PerformanceAnalysis}
R.~Jain, \emph{The art of computer systems performance analysis - techniques
  for experimental design, measurement, simulation, and modeling}, ser. Wiley
  professional computing.\hskip 1em plus 0.5em minus 0.4em\relax Wiley, 1991.

\bibitem{imagenet}
O.~Russakovsky, J.~Deng \emph{et~al.}, ``Imagenet large scale visual
  recognition challenge,'' 2015.

\bibitem{glue}
A.~Wang, A.~Singh \emph{et~al.}, ``Glue: A multi-task benchmark and analysis
  platform for natural language understanding,'' 2019.

\bibitem{2019arXiv191001500M}
P.~{Mattson}, C.~{Cheng} \emph{et~al.}, ``{MLPerf Training Benchmark},''
  \emph{arXiv e-prints}, p. arXiv:1910.01500, Oct. 2019.

\bibitem{perera2021chook}
D.~Perera, I.~Akpabio \emph{et~al.}, ``Chook -- a comprehensive suite for
  generating binary optimization problems with planted solutions,'' 2021.

\bibitem{supermarq}
\BIBentryALTinterwordspacing
T.~Tomesh, P.~Gokhale \emph{et~al.}, ``Supermarq: A scalable quantum benchmark
  suite,'' 2022. [Online]. Available: \url{https://arxiv.org/abs/2202.11045}
\BIBentrySTDinterwordspacing

\bibitem{gard2021classically}
B.~T. Gard and A.~M. Meier, ``A classically efficient quantum scalable
  fermi-hubbard benchmark,'' 2021.

\bibitem{dallairedemers2020application}
P.-L. Dallaire-Demers, M.~Stechly \emph{et~al.}, ``An application benchmark for
  fermionic quantum simulations,'' 2020.

\bibitem{Dongarra02thelinpack}
J.~J. Dongarra, P.~Luszczek, and A.~Petitet, ``The linpack benchmark: Past,
  present, and future,'' 2002.

\bibitem{spec}
\BIBentryALTinterwordspacing
{{SPEC}}. (2022) Standard performance evaluation corporation (spec):
  Benchmarks. [Online]. Available: \url{https://www.spec.org/benchmarks.html}
\BIBentrySTDinterwordspacing

\bibitem{RePEc:inm:orijoc:v:3:y:1991:i:4:p:376-384}
\BIBentryALTinterwordspacing
G.~Reinelt, ``{TSPLIB — A Traveling Salesman Problem Library},''
  \emph{INFORMS Journal on Computing}, vol.~3, no.~4, pp. 376--384, November
  1991. [Online]. Available:
  \url{https://ideas.repec.org/a/inm/orijoc/v3y1991i4p376-384.html}
\BIBentrySTDinterwordspacing

\bibitem{sat}
\BIBentryALTinterwordspacing
(2022) International sat competition. [Online]. Available:
  \url{http://www.satcompetition.org/}
\BIBentrySTDinterwordspacing

\bibitem{mccaskey2019quantum}
A.~J. McCaskey, Z.~P. Parks \emph{et~al.}, ``Quantum chemistry as a benchmark
  for near-term quantum computers,'' 2019.

\bibitem{Willsch}
M.~Willsch, D.~Willsch \emph{et~al.}, ``Benchmarking the quantum approximate
  optimization algorithm,'' \emph{Quantum Information Processing}, vol.~19, 06
  2020.

\bibitem{Katzgraber_PhysRevApplied.12.014004}
\BIBentryALTinterwordspacing
A.~Perdomo-Ortiz, A.~Feldman \emph{et~al.}, ``Readiness of quantum optimization
  machines for industrial applications,'' \emph{Phys. Rev. Applied}, vol.~12,
  p. 014004, Jul 2019. [Online]. Available:
  \url{https://link.aps.org/doi/10.1103/PhysRevApplied.12.014004}
\BIBentrySTDinterwordspacing

\bibitem{yarkoni2021multicar}
S.~Yarkoni, A.~Alekseyenko \emph{et~al.}, ``Multi-car paint shop optimization
  with quantum annealing,'' in \emph{2021 IEEE International Conference on
  Quantum Computing and Engineering (QCE)}, 2021, pp. 35--41.

\bibitem{accel}
G.~Juckeland, W.~Brantley \emph{et~al.}, ``Spec accel: A standard application
  suite for measuring hardware accelerator performance,'' 11 2014.

\bibitem{mpi}
M.~Müller, M.~van Waveren \emph{et~al.}, ``Spec mpi2007-an application
  benchmark suite for parallel systems using mpi,'' \emph{Concurrency and
  Computation: Practice and Experience}, vol.~22, pp. 191--205, 02 2010.

\bibitem{BlumeKohout2020volumetricframework}
\BIBentryALTinterwordspacing
R.~Blume-Kohout and K.~C. Young, ``A volumetric framework for quantum computer
  benchmarks,'' \emph{{Quantum}}, vol.~4, p. 362, Nov. 2020. [Online].
  Available: \url{https://doi.org/10.22331/q-2020-11-15-362}
\BIBentrySTDinterwordspacing

\bibitem{arline}
\BIBentryALTinterwordspacing
(2021) Standard performance evaluation corporation (spec): Benchmarks.
  [Online]. Available: \url{https://github.com/ArlineQ/arline_benchmarks}
\BIBentrySTDinterwordspacing

\bibitem{QASMBench}
\BIBentryALTinterwordspacing
A.~Li, S.~Stein \emph{et~al.}, ``Qasmbench: A low-level qasm benchmark suite
  for nisq evaluation and simulation,'' 2020. [Online]. Available:
  \url{https://arxiv.org/abs/2005.13018}
\BIBentrySTDinterwordspacing

\bibitem{shift_scheduling}
\BIBentryALTinterwordspacing
Schedulingbenchmarks.org. (2022) Shift scheduling benchmark data sets.
  [Online]. Available: \url{http://www.schedulingbenchmarks.org/other.html}
\BIBentrySTDinterwordspacing

\bibitem{clusterdata:Wilkes2020a}
J.~Wilkes, ``{Google} cluster-usage traces v3,'' Google Inc., Mountain View,
  CA, USA, Technical Report, Apr. 2020, posted at
  \url{https://github.com/google/cluster-data/blob/master/ClusterData2019.md}.

\bibitem{osti_1756458}
E.~Grant, T.~S. Humble, and B.~Stump, ``Benchmarking quantum annealing controls
  with portfolio optimization,'' \emph{Physical Review Applied}, vol.~15,
  no.~1, 1 2021.

\bibitem{Mills_2021}
\BIBentryALTinterwordspacing
D.~Mills, S.~Sivarajah \emph{et~al.}, ``Application-motivated, holistic
  benchmarking of a full quantum computing stack,'' \emph{Quantum}, vol.~5, p.
  415, Mar 2021. [Online]. Available:
  \url{http://dx.doi.org/10.22331/q-2021-03-22-415}
\BIBentrySTDinterwordspacing

\bibitem{9108020}
M.~{Muradi} and R.~{Wanka}, ``Sample-based motion planning for multi-robot
  systems,'' in \emph{2020 6th International Conference on Control, Automation
  and Robotics (ICCAR)}, 2020, pp. 130--138.

\bibitem{sat-ai}
\BIBentryALTinterwordspacing
N.~Narodytska, A.~Ignatiev \emph{et~al.}, ``Learning optimal decision trees
  with {SAT}.''\hskip 1em plus 0.5em minus 0.4em\relax International Joint
  Conferences on Artificial Intelligence Organization, Jul. 2018. [Online].
  Available: \url{https://doi.org/10.24963/ijcai.2018/189}
\BIBentrySTDinterwordspacing

\bibitem{sat-circuit-design}
T.~Hong, Y.~Li \emph{et~al.}, ``Qed: Quick error detection tests for effective
  post-silicon validation,'' in \emph{2010 IEEE International Test Conference},
  2010, pp. 1--10.

\bibitem{sat-bioinformatics}
I.~Lynce and J.~Marques-Silva, ``{SAT} in bioinformatics: Making the case with
  haplotype inference.''\hskip 1em plus 0.5em minus 0.4em\relax Springer Berlin
  Heidelberg, 2006, pp. 136--141.

\bibitem{satproceedings2021}
\BIBentryALTinterwordspacing
C.-M. Li and F.~Many{\`{a}}, Eds., \emph{Theory and Applications of
  Satisfiability Testing {\textendash} {SAT} 2021}.\hskip 1em plus 0.5em minus
  0.4em\relax Springer International Publishing, 2021. [Online]. Available:
  \url{https://doi.org/10.1007/978-3-030-80223-3}
\BIBentrySTDinterwordspacing

\bibitem{softhardcons}
\BIBentryALTinterwordspacing
L.~C. Min and M.~Felip, ``Maxsat, hard and soft constraints,'' \emph{Frontiers
  in Artificial Intelligence and Applications}, vol. 185, no. Handbook of
  Satisfiability, p. 613–631, 2009. [Online]. Available:
  \url{https://doi.org/10.3233/978-1-58603-929-5-613}
\BIBentrySTDinterwordspacing

\bibitem{skiena2008the}
S.~Skiena, \emph{The algorithm design manual}.\hskip 1em plus 0.5em minus
  0.4em\relax London: Springer, 2008.

\bibitem{tseitin}
G.~S. Tseitin, ``On the complexity of derivation in propositional calculus,''
  1983.

\bibitem{dinneen-qubo}
\BIBentryALTinterwordspacing
M.~J. Dinneen. (2016) Maximum 3-sat as qubo. [Online]. Available:
  \url{https://canvas.auckland.ac.nz/courses/14782/files/574983/download?verifier=1xqRikUjTEBwm8PnObD8YVmKdeEhZ9Ui8axW8HwP&wrap=1}
\BIBentrySTDinterwordspacing

\bibitem{choi_different_2011}
\BIBentryALTinterwordspacing
V.~Choi, ``Different {Adiabatic} {Quantum} {Optimization} {Algorithms} for the
  {NP}-{Complete} {Exact} {Cover} and {3SAT} {Problems},'' \emph{Proceedings of
  the National Academy of Sciences}, vol. 108, no.~7, pp. E19--E20, Feb. 2011,
  arXiv: 1010.1221. [Online]. Available: \url{http://arxiv.org/abs/1010.1221}
\BIBentrySTDinterwordspacing

\bibitem{DBLP:journals/corr/abs-1811-11538}
\BIBentryALTinterwordspacing
F.~W. Glover and G.~A. Kochenberger, ``A tutorial on formulating {QUBO}
  models,'' \emph{CoRR}, vol. abs/1811.11538, 2018. [Online]. Available:
  \url{http://arxiv.org/abs/1811.11538}
\BIBentrySTDinterwordspacing

\bibitem{zaman2021pyqubo}
M.~Zaman, K.~Tanahashi, and S.~Tanaka, ``Pyqubo: Python library for qubo
  creation,'' \emph{IEEE Transactions on Computers}, 2021.

\bibitem{ocean}
D.-W. Systems, ``D-wave ocean software documentation,''
  \url{https://docs.ocean.dwavesys.com/en/latest/index.html}, last accessed:
  Nov. 2021.

\bibitem{qiskit-optimize}
\BIBentryALTinterwordspacing
Q.~O. Developers. (2022) Qiskit optimization. [Online]. Available:
  \url{https://github.com/Qiskit/qiskit-optimization}
\BIBentrySTDinterwordspacing

\bibitem{farhi2014quantum}
E.~Farhi, J.~Goldstone, and S.~Gutmann, ``A quantum approximate optimization
  algorithm,'' 2014.

\bibitem{vqe_2014}
\BIBentryALTinterwordspacing
A.~Peruzzo, J.~McClean \emph{et~al.}, ``A variational eigenvalue solver on a
  photonic quantum processor,'' \emph{Nature Communications}, vol.~5, no.~1,
  Jul 2014. [Online]. Available: \url{http://dx.doi.org/10.1038/ncomms5213}
\BIBentrySTDinterwordspacing

\bibitem{bulger_implementing_2003}
\BIBentryALTinterwordspacing
D.~Bulger, W.~P. Baritompa, and G.~R. Wood, ``Implementing {Pure} {Adaptive}
  {Search} with {Grover}'s {Quantum} {Algorithm},'' \emph{Journal of
  Optimization Theory and Applications}, vol. 116, no.~3, pp. 517--529, Mar.
  2003. [Online]. Available: \url{https://doi.org/10.1023/A:1023061218864}
\BIBentrySTDinterwordspacing

\bibitem{qiskit}
G.~Aleksandrowicz, T.~Alexander \emph{et~al.}, ``Qiskit: An open-source
  framework for quantum computing,'' 2019.

\bibitem{bergholm2020pennylane}
V.~Bergholm, J.~Izaac \emph{et~al.}, ``Pennylane: Automatic differentiation of
  hybrid quantum-classical computations,'' 2020.

\bibitem{Wickham:2014:TidyData}
H.~Wickham \emph{et~al.}, ``Tidy data,'' \emph{Journal of Statistical
  Software}, vol.~59, no.~10, pp. 1--23, 2014.

\bibitem{simannealsampler}
\BIBentryALTinterwordspacing
{D-Wave Systems}, ``dwave-neal,'' D-Wave Systems, 2021. [Online]. Available:
  \url{https://docs.ocean.dwavesys.com/projects/neal/en/latest/reference/sampler.html}
\BIBentrySTDinterwordspacing

\bibitem{networkX}
N.~Developers,
  \url{https://networkx.org/documentation/stable/reference/algorithms/generated/networkx.algorithms.approximation.traveling_salesman.greedy_tsp.html},
  last accessed: Nov. 2021.

\bibitem{braket}
\BIBentryALTinterwordspacing
(2022) Amazon braket: Accelerate quantum computing research. [Online].
  Available: \url{https://aws.amazon.com/braket/}
\BIBentrySTDinterwordspacing

\bibitem{jones}
\BIBentryALTinterwordspacing
T.~Jones and J.~Gacon, ``Efficient calculation of gradients in classical
  simulations of variational quantum algorithms,'' 2020. [Online]. Available:
  \url{https://arxiv.org/abs/2009.02823}
\BIBentrySTDinterwordspacing

\bibitem{pennylaneGPU}
\BIBentryALTinterwordspacing
(2022) Lightning-fast simulations with pennylane and the nvidia cuquantum sdk.
  [Online]. Available:
  \url{https://pennylane.ai/blog/2022/07/lightning-fast-simulations-with-pennylane-and-the-nvidia-cuquantum-sdk/}
\BIBentrySTDinterwordspacing

\bibitem{Egger_2021}
\BIBentryALTinterwordspacing
D.~J. Egger, J.~Mare{\v{c} }ek, and S.~Woerner, ``Warm-starting quantum
  optimization,'' \emph{Quantum}, vol.~5, p. 479, jun 2021. [Online].
  Available: \url{https://doi.org/10.22331%2Fq-2021-06-17-479}
\BIBentrySTDinterwordspacing

\bibitem{qpack}
\BIBentryALTinterwordspacing
K.~Mesman, Z.~Al-Ars, and M.~M\"{o}ller, ``Qpack: Quantum approximate
  optimization algorithms as universal benchmark for quantum computers,'' 2021.
  [Online]. Available: \url{https://arxiv.org/abs/2103.17193}
\BIBentrySTDinterwordspacing

\bibitem{alteroper}
S.~Hadfield, Z.~Wang \emph{et~al.}, ``From the quantum approximate optimization
  algorithm to a quantum alternating operator ansatz,'' \emph{Algorithms},
  vol.~12, no.~2, 2019.

\bibitem{sat-solver-rc2}
A.~Ignatiev, A.~Morgado, and J.~Marques-Silva, ``Rc2: an efficient maxsat
  solver,'' \emph{Journal on Satisfiability, Boolean Modeling and Computation},
  vol.~11, pp. 53--64, 09 2019.

\bibitem{Lucas_2014}
\BIBentryALTinterwordspacing
A.~Lucas, ``Ising formulations of many np problems,'' \emph{Frontiers in
  Physics}, vol.~2, 2014. [Online]. Available:
  \url{http://dx.doi.org/10.3389/fphy.2014.00005}
\BIBentrySTDinterwordspacing

\bibitem{Joshi2018}
\BIBentryALTinterwordspacing
S.~Joshi, P.~Kumar \emph{et~al.}, ``Approximation strategies for incomplete
  {MaxSAT},'' in \emph{Lecture Notes in Computer Science}.\hskip 1em plus 0.5em
  minus 0.4em\relax Springer International Publishing, 2018, pp. 219--228.
  [Online]. Available: \url{https://doi.org/10.1007/978-3-319-98334-9_15}
\BIBentrySTDinterwordspacing

\bibitem{Bian2020}
\BIBentryALTinterwordspacing
Z.~Bian, F.~Chudak \emph{et~al.}, ``Solving {SAT} (and {MaxSAT}) with a quantum
  annealer: Foundations, encodings, and preliminary results,''
  \emph{Information and Computation}, vol. 275, p. 104609, Dec. 2020. [Online].
  Available: \url{https://doi.org/10.1016/j.ic.2020.104609}
\BIBentrySTDinterwordspacing

\bibitem{Chancellor2016}
\BIBentryALTinterwordspacing
N.~Chancellor, S.~Zohren \emph{et~al.}, ``A direct mapping of max k-{SAT} and
  high order parity checks to a chimera graph,'' \emph{Scientific Reports},
  vol.~6, no.~1, Nov. 2016. [Online]. Available:
  \url{https://doi.org/10.1038/srep37107}
\BIBentrySTDinterwordspacing

\bibitem{znidaric2006}
\BIBentryALTinterwordspacing
M.~{\v{Z}}nidari{\v{c}} and M.~Horvat, ``Exponential complexity of an adiabatic
  algorithm for an {NP}-complete problem,'' \emph{Physical Review A}, vol.~73,
  no.~2, Feb. 2006. [Online]. Available:
  \url{https://doi.org/10.1103/physreva.73.022329}
\BIBentrySTDinterwordspacing

\bibitem{Azinovic2017}
\BIBentryALTinterwordspacing
M.~Azinovi{\'{c}}, D.~Herr \emph{et~al.}, ``Assessment of quantum annealing for
  the construction of satisfiability filters,'' \emph{{SciPost} Physics},
  vol.~2, no.~2, Apr. 2017. [Online]. Available:
  \url{https://doi.org/10.21468/scipostphys.2.2.013}
\BIBentrySTDinterwordspacing

\end{thebibliography}



\begin{thebibliography}{51}
\ifx \bisbn   \undefined \def \bisbn  #1{ISBN #1}\fi
\ifx \binits  \undefined \def \binits#1{#1}\fi
\ifx \bauthor  \undefined \def \bauthor#1{#1}\fi
\ifx \batitle  \undefined \def \batitle#1{#1}\fi
\ifx \bjtitle  \undefined \def \bjtitle#1{#1}\fi
\ifx \bvolume  \undefined \def \bvolume#1{\textbf{#1}}\fi
\ifx \byear  \undefined \def \byear#1{#1}\fi
\ifx \bissue  \undefined \def \bissue#1{#1}\fi
\ifx \bfpage  \undefined \def \bfpage#1{#1}\fi
\ifx \blpage  \undefined \def \blpage #1{#1}\fi
\ifx \burl  \undefined \def \burl#1{\textsf{#1}}\fi
\ifx \doiurl  \undefined \def \doiurl#1{\textsf{#1}}\fi
\ifx \betal  \undefined \def \betal{\textit{et al.}}\fi
\ifx \binstitute  \undefined \def \binstitute#1{#1}\fi
\ifx \binstitutionaled  \undefined \def \binstitutionaled#1{#1}\fi
\ifx \bctitle  \undefined \def \bctitle#1{#1}\fi
\ifx \beditor  \undefined \def \beditor#1{#1}\fi
\ifx \bpublisher  \undefined \def \bpublisher#1{#1}\fi
\ifx \bbtitle  \undefined \def \bbtitle#1{#1}\fi
\ifx \bedition  \undefined \def \bedition#1{#1}\fi
\ifx \bseriesno  \undefined \def \bseriesno#1{#1}\fi
\ifx \blocation  \undefined \def \blocation#1{#1}\fi
\ifx \bsertitle  \undefined \def \bsertitle#1{#1}\fi
\ifx \bsnm \undefined \def \bsnm#1{#1}\fi
\ifx \bsuffix \undefined \def \bsuffix#1{#1}\fi
\ifx \bparticle \undefined \def \bparticle#1{#1}\fi
\ifx \barticle \undefined \def \barticle#1{#1}\fi
\ifx \bconfdate \undefined \def \bconfdate #1{#1}\fi
\ifx \botherref \undefined \def \botherref #1{#1}\fi
\ifx \url \undefined \def \url#1{\textsf{#1}}\fi
\ifx \bchapter \undefined \def \bchapter#1{#1}\fi
\ifx \bbook \undefined \def \bbook#1{#1}\fi
\ifx \bcomment \undefined \def \bcomment#1{#1}\fi
\ifx \oauthor \undefined \def \oauthor#1{#1}\fi
\ifx \citeauthoryear \undefined \def \citeauthoryear#1{#1}\fi
\ifx \endbibitem  \undefined \def \endbibitem {}\fi
\ifx \bconflocation  \undefined \def \bconflocation#1{#1}\fi
\ifx \arxivurl  \undefined \def \arxivurl#1{\textsf{#1}}\fi
\csname PreBibitemsHook\endcsname

\bibitem{digitaleWelt}
\begin{barticle}
\bauthor{\bsnm{Luckow}, \binits{A.}},
\bauthor{\bsnm{Klepsch}, \binits{J.}},
\bauthor{\bsnm{Pichlmeier}, \binits{J.}}:
\batitle{Quantum computing: Towards industry reference problems}.
\bjtitle{Digitale Welt}
\bvolume{2},
\bfpage{38}--\blpage{45}
(\byear{2021})
\end{barticle}
\endbibitem

\bibitem{qutac_epj}
\begin{botherref}
\oauthor{\bsnm{{{Quantum Technology and Application Consortium: Andreas
  Bayerstadler; Guillaume Becquin; Julia Binder; Thierry Botter; Hans Ehm;
  Thomas Ehmer; Marvin Erdmann; Norbert Gaus; Philipp Harbach; Maximilian Hess;
  Johannes Klepsch; Martin Leib; Sebastian Luber; Andre Luckow; Maximilian
  Mansky; Wolgang Mauerer; Florian Neukart; Christoph Niedermeier; Lilly
  Palackal; Carsten Polenz; Ruben Pfeiffer; Johanna Sepulveda; Tammo Sievers;
  Brian Standen; Michael Streif; Thomas Strohm; Clemens Utschig-Utschig; Daniel
  Volz; Horst Weiss; Fabian Winter}}}}:
Industry quantum applications
(vsl. 2022)
\end{botherref}
\endbibitem

\bibitem{PhysRevA.100.032328}
\begin{barticle}
\bauthor{\bsnm{Cross}, \binits{A.W.}},
\bauthor{\bsnm{Bishop}, \binits{L.S.}},
\bauthor{\bsnm{Sheldon}, \binits{S.}},
\bauthor{\bsnm{Nation}, \binits{P.D.}},
\bauthor{\bsnm{Gambetta}, \binits{J.M.}}:
\batitle{Validating quantum computers using randomized model circuits}.
\bjtitle{Phys. Rev. A}
\bvolume{100},
\bfpage{032328}
(\byear{2019}).
doi:\doiurl{10.1103/PhysRevA.100.032328}
\end{barticle}
\endbibitem

\bibitem{Bailey2011}
\begin{barticle}
\bauthor{\bsnm{Bailey}, \binits{D.H.}},
\bauthor{\bsnm{Barszcz}, \binits{E.}},
\bauthor{\bsnm{Barton}, \binits{J.T.}},
\bauthor{\bsnm{Browning}, \binits{D.S.}},
\bauthor{\bsnm{Carter}, \binits{R.L.}},
\bauthor{\bsnm{Dagum}, \binits{L.}},
\bauthor{\bsnm{Fatoohi}, \binits{R.A.}},
\bauthor{\bsnm{Frederickson}, \binits{P.O.}},
\bauthor{\bsnm{Lasinski}, \binits{T.A.}},
\bauthor{\bsnm{Schreiber}, \binits{R.S.}},
\bauthor{\bsnm{Simon}, \binits{H.D.}},
\bauthor{\bsnm{Venkatakrishnan}, \binits{V.}},
\bauthor{\bsnm{Weeratunga}, \binits{S.K.}}:
\batitle{The nas parallel benchmarks}.
\bjtitle{The International Journal of Supercomputing Applications}
\bvolume{5}(\bissue{3}),
\bfpage{63}--\blpage{73}
(\byear{1991}).
doi:\doiurl{10.1177/109434209100500306}.
\arxivurl{https://doi.org/10.1177/109434209100500306}
\end{barticle}
\endbibitem

\bibitem{qc_sim_list}
\begin{botherref}
List QC simulators.
\url{https://www.quantiki.org/wiki/list-qc-simulators}
(2021)
\end{botherref}
\endbibitem

\bibitem{De_Raedt_2019}
\begin{barticle}
\bauthor{\bsnm{De~Raedt}, \binits{H.}},
\bauthor{\bsnm{Jin}, \binits{F.}},
\bauthor{\bsnm{Willsch}, \binits{D.}},
\bauthor{\bsnm{Willsch}, \binits{M.}},
\bauthor{\bsnm{Yoshioka}, \binits{N.}},
\bauthor{\bsnm{Ito}, \binits{N.}},
\bauthor{\bsnm{Yuan}, \binits{S.}},
\bauthor{\bsnm{Michielsen}, \binits{K.}}:
\batitle{Massively parallel quantum computer simulator, eleven years later}.
\bjtitle{Computer Physics Communications}
\bvolume{237},
\bfpage{47}--\blpage{61}
(\byear{2019}).
doi:\doiurl{10.1016/j.cpc.2018.11.005}
\end{barticle}
\endbibitem

\bibitem{qulacs}
\begin{barticle}
\bauthor{\bsnm{Suzuki}, \binits{Y.}},
\bauthor{\bsnm{Kawase}, \binits{Y.}},
\bauthor{\bsnm{Masumura}, \binits{Y.}},
\bauthor{\bsnm{Hiraga}, \binits{Y.}},
\bauthor{\bsnm{Nakadai}, \binits{M.}},
\bauthor{\bsnm{Chen}, \binits{J.}},
\bauthor{\bsnm{Nakanishi}, \binits{K.M.}},
\bauthor{\bsnm{Mitarai}, \binits{K.}},
\bauthor{\bsnm{Imai}, \binits{R.}},
\bauthor{\bsnm{Tamiya}, \binits{S.}},
\bauthor{\bparticle{et} \bsnm{al.}}:
\batitle{Qulacs: a fast and versatile quantum circuit simulator for research
  purpose}.
\bjtitle{Quantum}
\bvolume{5},
\bfpage{559}
(\byear{2021}).
doi:\doiurl{10.22331/q-2021-10-06-559}
\end{barticle}
\endbibitem

\bibitem{willsch2021gpuaccelerated}
\begin{botherref}
\oauthor{\bsnm{Willsch}, \binits{D.}},
\oauthor{\bsnm{Willsch}, \binits{M.}},
\oauthor{\bsnm{Jin}, \binits{F.}},
\oauthor{\bsnm{Michielsen}, \binits{K.}},
\oauthor{\bsnm{Raedt}, \binits{H.D.}}:
GPU-accelerated simulations of quantum annealing and the quantum approximate
  optimization algorithm
(2021).
\arxivurl{2104.03293}
\end{botherref}
\endbibitem

\bibitem{cirac2021matrix}
\begin{botherref}
\oauthor{\bsnm{Cirac}, \binits{I.}},
\oauthor{\bsnm{Perez-Garcia}, \binits{D.}},
\oauthor{\bsnm{Schuch}, \binits{N.}},
\oauthor{\bsnm{Verstraete}, \binits{F.}}:
Matrix Product States and Projected Entangled Pair States: Concepts,
  Symmetries, and Theorems
(2021).
\arxivurl{2011.12127}
\end{botherref}
\endbibitem

\bibitem{Markov_2008}
\begin{barticle}
\bauthor{\bsnm{Markov}, \binits{I.L.}},
\bauthor{\bsnm{Shi}, \binits{Y.}}:
\batitle{Simulating quantum computation by contracting tensor networks}.
\bjtitle{SIAM Journal on Computing}
\bvolume{38}(\bissue{3}),
\bfpage{963}--\blpage{981}
(\byear{2008}).
doi:\doiurl{10.1137/050644756}
\end{barticle}
\endbibitem

\bibitem{Vidal_2003}
\begin{botherref}
\oauthor{\bsnm{Vidal}, \binits{G.}}:
Efficient classical simulation of slightly entangled quantum computations.
Physical Review Letters
\textbf{91}(14)
(2003).
doi:\doiurl{10.1103/physrevlett.91.147902}
\end{botherref}
\endbibitem

\bibitem{pan2021simulating}
\begin{botherref}
\oauthor{\bsnm{Pan}, \binits{F.}},
\oauthor{\bsnm{Zhang}, \binits{P.}}:
Simulating the Sycamore quantum supremacy circuits
(2021).
\arxivurl{2103.03074}
\end{botherref}
\endbibitem

\bibitem{Medvidovi__2021}
\begin{botherref}
\oauthor{\bsnm{Medvidović}, \binits{M.}},
\oauthor{\bsnm{Carleo}, \binits{G.}}:
Classical variational simulation of the quantum approximate optimization
  algorithm.
npj Quantum Information
\textbf{7}(1)
(2021).
doi:\doiurl{10.1038/s41534-021-00440-z}
\end{botherref}
\endbibitem

\bibitem{10.5555/1972505}
\begin{bbook}
\bauthor{\bsnm{Nielsen}, \binits{M.A.}},
\bauthor{\bsnm{Chuang}, \binits{I.L.}}:
\bbtitle{Quantum Computation and Quantum Information: 10th Anniversary
  Edition},
\bedition{10th} edn.
\bpublisher{Cambridge University Press},
\blocation{USA}
(\byear{2011})
\end{bbook}
\endbibitem

\bibitem{li2020density}
\begin{bchapter}
\bauthor{\bsnm{Li}, \binits{A.}},
\bauthor{\bsnm{Subasi}, \binits{O.}},
\bauthor{\bsnm{Yang}, \binits{X.}},
\bauthor{\bsnm{Krishnamoorthy}, \binits{S.}}:
\bctitle{Density matrix quantum circuit simulation via the bsp machine on
  modern gpu clusters}.
In: \bbtitle{Proceedings of the International Conference for High Performance
  Computing, Networking, Storage and Analysis}
(\byear{2020})
\end{bchapter}
\endbibitem

\bibitem{qiskit}
\begin{botherref}
\oauthor{\bsnm{Aleksandrowicz}, \binits{G.}},
\oauthor{\bsnm{Alexander}, \binits{T.}},
\oauthor{\bsnm{Barkoutsos}, \binits{P.}},
\oauthor{\bsnm{Bello}, \binits{L.}},
\oauthor{\bsnm{Ben-Haim}, \binits{Y.}},
\oauthor{\bsnm{Bucher}, \binits{D.}},
\oauthor{\bsnm{Cabrera-Hernandez}, \binits{F.J.}},
\oauthor{\bsnm{Carballo-Franquis}, \binits{J.}},
\oauthor{\bsnm{Chen}, \binits{A.}},
\oauthor{\bsnm{Chen}, \binits{C.-F.}}:
Qiskit: An Open-source Framework for Quantum Computing
(2019).
doi:\doiurl{10.5281/zenodo.2562110}
\end{botherref}
\endbibitem

\bibitem{circ}
\begin{botherref}
Circ: A Python framework for creating, editing, and invoking Noisy Intermediate
  Scale Quantum (NISQ) circuits.
\url{https://github.com/quantumlib/Cirq}
(2021)
\end{botherref}
\endbibitem

\bibitem{braket}
\begin{botherref}
Amazon Braket Python SDK: A python SDK for interacting with quantum devices on
  Amazon Braket.
\url{https://github.com/aws/amazon-braket-sdk-python}
(2021)
\end{botherref}
\endbibitem

\bibitem{ibm}
\begin{botherref}
IBM Quantum.
\url{https://quantum-computing.ibm.com/}
(2021)
\end{botherref}
\endbibitem

\bibitem{google}
\begin{botherref}
\oauthor{\bsnm{Google}}:
Quantum Computer Datasheet.
\url{https://quantumai.google/hardware/datasheet/weber.pdf}
(2021)
\end{botherref}
\endbibitem

\bibitem{rigetti}
\begin{botherref}
Rigetti.
\url{https://www.rigetti.com/get-quantum}
(2021)
\end{botherref}
\endbibitem

\bibitem{ionq}
\begin{botherref}
IonQ.
\url{https://ionq.com/}
(2021)
\end{botherref}
\endbibitem

\bibitem{honeywell}
\begin{botherref}
Honeywell.
\url{https://www.honeywell.com/us/en/company/quantum}
(2021)
\end{botherref}
\endbibitem

\bibitem{dwave}
\begin{botherref}
D-Wave: Leap.
\url{https://cloud.dwavesys.com/}
(2021)
\end{botherref}
\endbibitem

\bibitem{martiel2021benchmarking}
\begin{botherref}
\oauthor{\bsnm{Martiel}, \binits{S.}},
\oauthor{\bsnm{Ayral}, \binits{T.}},
\oauthor{\bsnm{Allouche}, \binits{C.}}:
Benchmarking quantum co-processors in an application-centric, hardware-agnostic
  and scalable way
(2021).
\arxivurl{2102.12973}
\end{botherref}
\endbibitem

\bibitem{lubinski2021applicationoriented}
\begin{botherref}
\oauthor{\bsnm{Lubinski}, \binits{T.}},
\oauthor{\bsnm{Johri}, \binits{S.}},
\oauthor{\bsnm{Varosy}, \binits{P.}},
\oauthor{\bsnm{Coleman}, \binits{J.}},
\oauthor{\bsnm{Zhao}, \binits{L.}},
\oauthor{\bsnm{Necaise}, \binits{J.}},
\oauthor{\bsnm{Baldwin}, \binits{C.H.}},
\oauthor{\bsnm{Mayer}, \binits{K.}},
\oauthor{\bsnm{Proctor}, \binits{T.}}:
Application-Oriented Performance Benchmarks for Quantum Computing
(2021).
\arxivurl{2110.03137}
\end{botherref}
\endbibitem

\bibitem{Dongarra02thelinpack}
\begin{botherref}
\oauthor{\bsnm{Dongarra}, \binits{J.J.}},
\oauthor{\bsnm{Luszczek}, \binits{P.}},
\oauthor{\bsnm{Petitet}, \binits{A.}}:
The LINPACK benchmark: Past, present, and future
(2002)
\end{botherref}
\endbibitem

\bibitem{doi:10.1177/109434209100500306}
\begin{barticle}
\bauthor{\bsnm{Bailey}, \binits{D.H.}},
\bauthor{\bsnm{Barszcz}, \binits{E.}},
\bauthor{\bsnm{Barton}, \binits{J.T.}},
\bauthor{\bsnm{Browning}, \binits{D.S.}},
\bauthor{\bsnm{Carter}, \binits{R.L.}},
\bauthor{\bsnm{Dagum}, \binits{L.}},
\bauthor{\bsnm{Fatoohi}, \binits{R.A.}},
\bauthor{\bsnm{Frederickson}, \binits{P.O.}},
\bauthor{\bsnm{Lasinski}, \binits{T.A.}},
\bauthor{\bsnm{Schreiber}, \binits{R.S.}},
\bauthor{\bsnm{Simon}, \binits{H.D.}},
\bauthor{\bsnm{Venkatakrishnan}, \binits{V.}},
\bauthor{\bsnm{Weeratunga}, \binits{S.K.}}:
\batitle{The nas parallel benchmarks}.
\bjtitle{The International Journal of Supercomputing Applications}
\bvolume{5}(\bissue{3}),
\bfpage{63}--\blpage{73}
(\byear{1991}).
doi:\doiurl{10.1177/109434209100500306}.
\arxivurl{https://doi.org/10.1177/109434209100500306}
\end{barticle}
\endbibitem

\bibitem{imagenet}
\begin{botherref}
\oauthor{\bsnm{Russakovsky}, \binits{O.}},
\oauthor{\bsnm{Deng}, \binits{J.}},
\oauthor{\bsnm{Su}, \binits{H.}},
\oauthor{\bsnm{Krause}, \binits{J.}},
\oauthor{\bsnm{Satheesh}, \binits{S.}},
\oauthor{\bsnm{Ma}, \binits{S.}},
\oauthor{\bsnm{Huang}, \binits{Z.}},
\oauthor{\bsnm{Karpathy}, \binits{A.}},
\oauthor{\bsnm{Khosla}, \binits{A.}},
\oauthor{\bsnm{Bernstein}, \binits{M.}},
\oauthor{\bsnm{Berg}, \binits{A.C.}},
\oauthor{\bsnm{Fei-Fei}, \binits{L.}}:
ImageNet Large Scale Visual Recognition Challenge
(2015).
\arxivurl{1409.0575}
\end{botherref}
\endbibitem

\bibitem{glue}
\begin{botherref}
\oauthor{\bsnm{Wang}, \binits{A.}},
\oauthor{\bsnm{Singh}, \binits{A.}},
\oauthor{\bsnm{Michael}, \binits{J.}},
\oauthor{\bsnm{Hill}, \binits{F.}},
\oauthor{\bsnm{Levy}, \binits{O.}},
\oauthor{\bsnm{Bowman}, \binits{S.R.}}:
GLUE: A Multi-Task Benchmark and Analysis Platform for Natural Language
  Understanding
(2019).
\arxivurl{1804.07461}
\end{botherref}
\endbibitem

\bibitem{mlcommons}
\begin{botherref}
MLCommons.
\url{https://mlcommons.org/en/}
(2021)
\end{botherref}
\endbibitem

\bibitem{2019arXiv191001500M}
\begin{botherref}
\oauthor{\bsnm{{Mattson}}, \binits{P.}},
\oauthor{\bsnm{{Cheng}}, \binits{C.}},
\oauthor{\bsnm{{Coleman}}, \binits{C.}},
\oauthor{\bsnm{{Diamos}}, \binits{G.}},
\oauthor{\bsnm{{Micikevicius}}, \binits{P.}},
\oauthor{\bsnm{{Patterson}}, \binits{D.}},
\oauthor{\bsnm{{Tang}}, \binits{H.}},
\oauthor{\bsnm{{Wei}}, \binits{G.-Y.}},
\oauthor{\bsnm{{Bailis}}, \binits{P.}},
\oauthor{\bsnm{{Bittorf}}, \binits{V.}},
\oauthor{\bsnm{{Brooks}}, \binits{D.}},
\oauthor{\bsnm{{Chen}}, \binits{D.}},
\oauthor{\bsnm{{Dutta}}, \binits{D.}},
\oauthor{\bsnm{{Gupta}}, \binits{U.}},
\oauthor{\bsnm{{Hazelwood}}, \binits{K.}},
\oauthor{\bsnm{{Hock}}, \binits{A.}},
\oauthor{\bsnm{{Huang}}, \binits{X.}},
\oauthor{\bsnm{{Ike}}, \binits{A.}},
\oauthor{\bsnm{{Jia}}, \binits{B.}},
\oauthor{\bsnm{{Kang}}, \binits{D.}},
\oauthor{\bsnm{{Kanter}}, \binits{D.}},
\oauthor{\bsnm{{Kumar}}, \binits{N.}},
\oauthor{\bsnm{{Liao}}, \binits{J.}},
\oauthor{\bsnm{{Ma}}, \binits{G.}},
\oauthor{\bsnm{{Narayanan}}, \binits{D.}},
\oauthor{\bsnm{{Oguntebi}}, \binits{T.}},
\oauthor{\bsnm{{Pekhimenko}}, \binits{G.}},
\oauthor{\bsnm{{Pentecost}}, \binits{L.}},
\oauthor{\bsnm{{Janapa Reddi}}, \binits{V.}},
\oauthor{\bsnm{{Robie}}, \binits{T.}},
\oauthor{\bsnm{{St. John}}, \binits{T.}},
\oauthor{\bsnm{{Tabaru}}, \binits{T.}},
\oauthor{\bsnm{{Wu}}, \binits{C.-J.}},
\oauthor{\bsnm{{Xu}}, \binits{L.}},
\oauthor{\bsnm{{Yamazaki}}, \binits{M.}},
\oauthor{\bsnm{{Young}}, \binits{C.}},
\oauthor{\bsnm{{Zaharia}}, \binits{M.}}:
{MLPerf Training Benchmark}.
arXiv e-prints,
1910--01500
(2019).
\arxivurl{1910.01500}
\end{botherref}
\endbibitem

\bibitem{sat}
\begin{botherref}
International SAT Competition.
\url{http://www.satcompetition.org/}
(2021)
\end{botherref}
\endbibitem

\bibitem{shift_scheduling}
\begin{botherref}
Shift Scheduling Benchmark Data Sets.
\url{http://www.schedulingbenchmarks.org/other.html}
\end{botherref}
\endbibitem

\bibitem{clusterdata:Wilkes2020a}
\begin{botherref}
\oauthor{\bsnm{Wilkes}, \binits{J.}}:
{Google} cluster-usage traces v3.
Technical report,
Google Inc.,
Mountain View, CA, USA
(April 2020).
Posted at
  \url{https://github.com/google/cluster-data/blob/master/ClusterData2019.md}
\end{botherref}
\endbibitem

\bibitem{RePEc:inm:orijoc:v:3:y:1991:i:4:p:376-384}
\begin{barticle}
\bauthor{\bsnm{Reinelt}, \binits{G.}}:
\batitle{{TSPLIB — A Traveling Salesman Problem Library}}.
\bjtitle{INFORMS Journal on Computing}
\bvolume{3}(\bissue{4}),
\bfpage{376}--\blpage{384}
(\byear{1991}).
doi:\doiurl{10.1287/ijoc.3.4.376}
\end{barticle}
\endbibitem

\bibitem{TSPLIB95}
\begin{botherref}
\oauthor{\bsnm{Reinelt}, \binits{G.}}:
TSPLIB 95.
\url{http://comopt.ifi.uni-heidelberg.de/software/TSPLIB95/tsp95.pdf}
(last accessed 2021)
\end{botherref}
\endbibitem

\bibitem{BlumeKohout2020volumetricframework}
\begin{barticle}
\bauthor{\bsnm{Blume-Kohout}, \binits{R.}},
\bauthor{\bsnm{Young}, \binits{K.C.}}:
\batitle{A volumetric framework for quantum computer benchmarks}.
\bjtitle{{Quantum}}
\bvolume{4},
\bfpage{362}
(\byear{2020}).
doi:\doiurl{10.22331/q-2020-11-15-362}
\end{barticle}
\endbibitem

\bibitem{wack2021quality}
\begin{botherref}
\oauthor{\bsnm{Wack}, \binits{A.}},
\oauthor{\bsnm{Paik}, \binits{H.}},
\oauthor{\bsnm{Javadi-Abhari}, \binits{A.}},
\oauthor{\bsnm{Jurcevic}, \binits{P.}},
\oauthor{\bsnm{Faro}, \binits{I.}},
\oauthor{\bsnm{Gambetta}, \binits{J.M.}},
\oauthor{\bsnm{Johnson}, \binits{B.R.}}:
Quality, Speed, and Scale: three key attributes to measure the performance of
  near-term quantum computers
(2021).
\arxivurl{2110.14108}
\end{botherref}
\endbibitem

\bibitem{osti_1756458}
\begin{botherref}
\oauthor{\bsnm{Grant}, \binits{E.}},
\oauthor{\bsnm{Humble}, \binits{T.S.}},
\oauthor{\bsnm{Stump}, \binits{B.}}:
Benchmarking quantum annealing controls with portfolio optimization.
Physical Review Applied
\textbf{15}(1)
(2021).
doi:\doiurl{10.1103/physrevapplied.15.014012}
\end{botherref}
\endbibitem

\bibitem{osti_1759113}
\begin{barticle}
\bauthor{\bsnm{Pang}, \binits{Y.}},
\bauthor{\bsnm{Coffrin}, \binits{C.}},
\bauthor{\bsnm{Lokhov}, \binits{A.Y.}},
\bauthor{\bsnm{Vuffray}, \binits{M.}}:
\batitle{The potential of quantum annealing for rapid solution structure
  identification}.
\bjtitle{Constraints}
(\byear{2020}).
doi:\doiurl{10.1007/s10601-020-09315-0}
\end{barticle}
\endbibitem

\bibitem{Katzgraber_PhysRevApplied.12.014004}
\begin{barticle}
\bauthor{\bsnm{Perdomo-Ortiz}, \binits{A.}},
\bauthor{\bsnm{Feldman}, \binits{A.}},
\bauthor{\bsnm{Ozaeta}, \binits{A.}},
\bauthor{\bsnm{Isakov}, \binits{S.V.}},
\bauthor{\bsnm{Zhu}, \binits{Z.}},
\bauthor{\bsnm{O'Gorman}, \binits{B.}},
\bauthor{\bsnm{Katzgraber}, \binits{H.G.}},
\bauthor{\bsnm{Diedrich}, \binits{A.}},
\bauthor{\bsnm{Neven}, \binits{H.}},
\bauthor{\bparticle{de} \bsnm{Kleer}, \binits{J.}},
\bauthor{\bsnm{Lackey}, \binits{B.}},
\bauthor{\bsnm{Biswas}, \binits{R.}}:
\batitle{Readiness of quantum optimization machines for industrial
  applications}.
\bjtitle{Phys. Rev. Applied}
\bvolume{12},
\bfpage{014004}
(\byear{2019}).
doi:\doiurl{10.1103/PhysRevApplied.12.014004}
\end{barticle}
\endbibitem

\bibitem{yarkoni2021multicar}
\begin{botherref}
\oauthor{\bsnm{Yarkoni}, \binits{S.}},
\oauthor{\bsnm{Alekseyenko}, \binits{A.}},
\oauthor{\bsnm{Streif}, \binits{M.}},
\oauthor{\bsnm{Dollen}, \binits{D.V.}},
\oauthor{\bsnm{Neukart}, \binits{F.}},
\oauthor{\bsnm{Bäck}, \binits{T.}}:
Multi-car paint shop optimization with quantum annealing
(2021).
\arxivurl{2109.07876}
\end{botherref}
\endbibitem

\bibitem{Willsch}
\begin{botherref}
\oauthor{\bsnm{Willsch}, \binits{M.}},
\oauthor{\bsnm{Willsch}, \binits{D.}},
\oauthor{\bsnm{Raedt}, \binits{H.}},
\oauthor{\bsnm{Michielsen}, \binits{K.}}:
Benchmarking the quantum approximate optimization algorithm.
Quantum Information Processing
\textbf{19}
(2020).
doi:\doiurl{10.1007/s11128-020-02692-8}
\end{botherref}
\endbibitem

\bibitem{mccaskey2019quantum}
\begin{botherref}
\oauthor{\bsnm{McCaskey}, \binits{A.J.}},
\oauthor{\bsnm{Parks}, \binits{Z.P.}},
\oauthor{\bsnm{Jakowski}, \binits{J.}},
\oauthor{\bsnm{Moore}, \binits{S.V.}},
\oauthor{\bsnm{Morris}, \binits{T.}},
\oauthor{\bsnm{Humble}, \binits{T.S.}},
\oauthor{\bsnm{Pooser}, \binits{R.C.}}:
Quantum Chemistry as a Benchmark for Near-Term Quantum Computers
(2019).
\arxivurl{1905.01534}
\end{botherref}
\endbibitem

\bibitem{dallairedemers2020application}
\begin{botherref}
\oauthor{\bsnm{Dallaire-Demers}, \binits{P.-L.}},
\oauthor{\bsnm{Stechly}, \binits{M.}},
\oauthor{\bsnm{Gonthier}, \binits{J.F.}},
\oauthor{\bsnm{Bashige}, \binits{N.T.}},
\oauthor{\bsnm{Romero}, \binits{J.}},
\oauthor{\bsnm{Cao}, \binits{Y.}}:
An application benchmark for fermionic quantum simulations
(2020).
\arxivurl{2003.01862}
\end{botherref}
\endbibitem

\bibitem{Mills_2021}
\begin{barticle}
\bauthor{\bsnm{Mills}, \binits{D.}},
\bauthor{\bsnm{Sivarajah}, \binits{S.}},
\bauthor{\bsnm{Scholten}, \binits{T.L.}},
\bauthor{\bsnm{Duncan}, \binits{R.}}:
\batitle{Application-motivated, holistic benchmarking of a full quantum
  computing stack}.
\bjtitle{Quantum}
\bvolume{5},
\bfpage{415}
(\byear{2021}).
doi:\doiurl{10.22331/q-2021-03-22-415}
\end{barticle}
\endbibitem

\bibitem{qutac}
\begin{botherref}
\oauthor{\bsnm{Klepsch}, \binits{J.}},
\oauthor{\bsnm{Kopp}, \binits{J.}},
\oauthor{\bsnm{Luckow}, \binits{A.}},
\oauthor{\bsnm{Weiss}, \binits{H.}},
\oauthor{\bsnm{Standen}, \binits{B.}},
\oauthor{\bsnm{Vozl}, \binits{D.}},
\oauthor{\bsnm{Utschig-Utschig}, \binits{C.}},
\oauthor{\bsnm{Streif}, \binits{M.}},
\oauthor{\bsnm{Strohm}, \binits{T.}},
\oauthor{\bsnm{Ehm}, \binits{H.}},
\oauthor{\bsnm{Luber}, \binits{S.}},
\oauthor{\bsnm{Richter}, \binits{J.}},
\oauthor{\bsnm{Zeyen}, \binits{B.}},
\oauthor{\bsnm{Harbach}, \binits{P.}},
\oauthor{\bsnm{Ehmer}, \binits{T.}},
\oauthor{\bsnm{Bayerstadler}, \binits{A.}},
\oauthor{\bsnm{Winter}, \binits{F.}},
\oauthor{\bsnm{Becquin}, \binits{G.}},
\oauthor{\bsnm{Polenz}, \binits{C.}},
\oauthor{\bsnm{Mauerer}, \binits{W.}},
\oauthor{\bsnm{Niedermeier}, \binits{C.}},
\oauthor{\bsnm{Gaus}, \binits{N.}},
\oauthor{\bsnm{Leib}, \binits{M.}},
\oauthor{\bsnm{Neukart}, \binits{F.}},
\oauthor{\bsnm{Botter}, \binits{T.}},
\oauthor{\bsnm{Sepulveda}, \binits{J.}},
\oauthor{\bsnm{Dombrowski}, \binits{E.}},
\oauthor{\bsnm{Leonhardt}, \binits{T.}},
\oauthor{\bsnm{Vornehm}, \binits{N.}}:
Industry Quantum Applications.
\url{https://www.qutac.de/wp-content/uploads/2021/07/QUTAC_Paper.pdf}
(2021)
\end{botherref}
\endbibitem

\bibitem{ferrari1978computer}
\begin{bbook}
\bauthor{\bsnm{Ferrari}, \binits{D.}}:
\bbtitle{Computer Systems Performance Evaluation}.
\bpublisher{Prentice-Hall}, \blocation{???}
(\byear{1978}).
\burl{https://books.google.de/books?id=geBQAAAAMAAJ}
\end{bbook}
\endbibitem

\bibitem{Jain:1991:PerformanceAnalysis}
\begin{bbook}
\bauthor{\bsnm{Jain}, \binits{R.}}:
\bbtitle{The Art of Computer Systems Performance Analysis - Techniques for
  Experimental Design, Measurement, Simulation, and Modeling}.
\bsertitle{Wiley professional computing}.
\bpublisher{Wiley}, \blocation{???}
(\byear{1991})
\end{bbook}
\endbibitem

\bibitem{Wickham:2014:TidyData}
\begin{barticle}
\bauthor{\bsnm{Wickham}, \binits{H.}}, \betal:
\batitle{Tidy data}.
\bjtitle{Journal of Statistical Software}
\bvolume{59}(\bissue{10}),
\bfpage{1}--\blpage{23}
(\byear{2014})
\end{barticle}
\endbibitem

\end{thebibliography}

\newcommand{\BMCxmlcomment}[1]{}

\BMCxmlcomment{

<refgrp>

<bibl id="B1">
  <title><p>Quantum Computing: Towards Industry Reference Problems</p></title>
  <aug>
    <au><snm>Luckow</snm><fnm>A</fnm></au>
    <au><snm>Klepsch</snm><fnm>J</fnm></au>
    <au><snm>Pichlmeier</snm><fnm>J</fnm></au>
  </aug>
  <source>Digitale Welt</source>
  <pubdate>2021</pubdate>
  <volume>2</volume>
  <fpage>38</fpage>
  <lpage>-45</lpage>
</bibl>

<bibl id="B2">
  <title><p>Industry Quantum Applications</p></title>
  <aug>
    <au><cnm>{{Quantum Technology and Application Consortium: Andreas
  Bayerstadler; Guillaume Becquin; Julia Binder; Thierry Botter; Hans Ehm;
  Thomas Ehmer; Marvin Erdmann; Norbert Gaus; Philipp Harbach; Maximilian Hess;
  Johannes Klepsch; Martin Leib; Sebastian Luber; Andre Luckow; Maximilian
  Mansky; Wolgang Mauerer; Florian Neukart; Christoph Niedermeier; Lilly
  Palackal; Carsten Polenz; Ruben Pfeiffer; Johanna Sepulveda; Tammo Sievers;
  Brian Standen; Michael Streif; Thomas Strohm; Clemens Utschig-Utschig; Daniel
  Volz; Horst Weiss; Fabian Winter}}</cnm></au>
  </aug>
  <source>submitted to EPJ Qunatum Technology, Springer</source>
  <pubdate>vsl. 2022</pubdate>
</bibl>

<bibl id="B3">
  <title><p>Validating quantum computers using randomized model
  circuits</p></title>
  <aug>
    <au><snm>Cross</snm><fnm>AW</fnm></au>
    <au><snm>Bishop</snm><fnm>LS</fnm></au>
    <au><snm>Sheldon</snm><fnm>S</fnm></au>
    <au><snm>Nation</snm><fnm>PD</fnm></au>
    <au><snm>Gambetta</snm><fnm>JM</fnm></au>
  </aug>
  <source>Phys. Rev. A</source>
  <publisher>American Physical Society</publisher>
  <pubdate>2019</pubdate>
  <volume>100</volume>
  <fpage>032328</fpage>
  <url>https://link.aps.org/doi/10.1103/PhysRevA.100.032328</url>
</bibl>

<bibl id="B4">
  <title><p>The Nas Parallel Benchmarks</p></title>
  <aug>
    <au><snm>Bailey</snm><fnm>D.H.</fnm></au>
    <au><snm>Barszcz</snm><fnm>E.</fnm></au>
    <au><snm>Barton</snm><fnm>J.T.</fnm></au>
    <au><snm>Browning</snm><fnm>D.S.</fnm></au>
    <au><snm>Carter</snm><fnm>R.L.</fnm></au>
    <au><snm>Dagum</snm><fnm>L.</fnm></au>
    <au><snm>Fatoohi</snm><fnm>R.A.</fnm></au>
    <au><snm>Frederickson</snm><fnm>P.O.</fnm></au>
    <au><snm>Lasinski</snm><fnm>T.A.</fnm></au>
    <au><snm>Schreiber</snm><fnm>R.S.</fnm></au>
    <au><snm>Simon</snm><fnm>H.D.</fnm></au>
    <au><snm>Venkatakrishnan</snm><fnm>V.</fnm></au>
    <au><snm>Weeratunga</snm><fnm>S.K.</fnm></au>
  </aug>
  <source>The International Journal of Supercomputing Applications</source>
  <pubdate>1991</pubdate>
  <volume>5</volume>
  <issue>3</issue>
  <fpage>63</fpage>
  <lpage>73</lpage>
  <url>https://doi.org/10.1177/109434209100500306</url>
</bibl>

<bibl id="B5">
  <title><p>List QC simulators</p></title>
  <source>\url{https://www.quantiki.org/wiki/list-qc-simulators}</source>
  <pubdate>2021</pubdate>
</bibl>

<bibl id="B6">
  <title><p>Massively parallel quantum computer simulator, eleven years
  later</p></title>
  <aug>
    <au><snm>De Raedt</snm><fnm>H</fnm></au>
    <au><snm>Jin</snm><fnm>F</fnm></au>
    <au><snm>Willsch</snm><fnm>D</fnm></au>
    <au><snm>Willsch</snm><fnm>M</fnm></au>
    <au><snm>Yoshioka</snm><fnm>N</fnm></au>
    <au><snm>Ito</snm><fnm>N</fnm></au>
    <au><snm>Yuan</snm><fnm>S</fnm></au>
    <au><snm>Michielsen</snm><fnm>K</fnm></au>
  </aug>
  <source>Computer Physics Communications</source>
  <publisher>Elsevier BV</publisher>
  <pubdate>2019</pubdate>
  <volume>237</volume>
  <fpage>47–61</fpage>
  <url>http://dx.doi.org/10.1016/j.cpc.2018.11.005</url>
</bibl>

<bibl id="B7">
  <title><p>Qulacs: a fast and versatile quantum circuit simulator for research
  purpose</p></title>
  <aug>
    <au><snm>Suzuki</snm><fnm>Y</fnm></au>
    <au><snm>Kawase</snm><fnm>Y</fnm></au>
    <au><snm>Masumura</snm><fnm>Y</fnm></au>
    <au><snm>Hiraga</snm><fnm>Y</fnm></au>
    <au><snm>Nakadai</snm><fnm>M</fnm></au>
    <au><snm>Chen</snm><fnm>J</fnm></au>
    <au><snm>Nakanishi</snm><fnm>KM</fnm></au>
    <au><snm>Mitarai</snm><fnm>K</fnm></au>
    <au><snm>Imai</snm><fnm>R</fnm></au>
    <au><snm>Tamiya</snm><fnm>S</fnm></au>
    <au><cnm>al.</cnm></au>
  </aug>
  <source>Quantum</source>
  <publisher>Verein zur Forderung des Open Access Publizierens in den
  Quantenwissenschaften</publisher>
  <pubdate>2021</pubdate>
  <volume>5</volume>
  <fpage>559</fpage>
  <url>http://dx.doi.org/10.22331/q-2021-10-06-559</url>
</bibl>

<bibl id="B8">
  <title><p>GPU-accelerated simulations of quantum annealing and the quantum
  approximate optimization algorithm</p></title>
  <aug>
    <au><snm>Willsch</snm><fnm>D</fnm></au>
    <au><snm>Willsch</snm><fnm>M</fnm></au>
    <au><snm>Jin</snm><fnm>F</fnm></au>
    <au><snm>Michielsen</snm><fnm>K</fnm></au>
    <au><snm>Raedt</snm><fnm>HD</fnm></au>
  </aug>
  <pubdate>2021</pubdate>
</bibl>

<bibl id="B9">
  <title><p>Matrix Product States and Projected Entangled Pair States:
  Concepts, Symmetries, and Theorems</p></title>
  <aug>
    <au><snm>Cirac</snm><fnm>I</fnm></au>
    <au><snm>Perez Garcia</snm><fnm>D</fnm></au>
    <au><snm>Schuch</snm><fnm>N</fnm></au>
    <au><snm>Verstraete</snm><fnm>F</fnm></au>
  </aug>
  <pubdate>2021</pubdate>
</bibl>

<bibl id="B10">
  <title><p>Simulating Quantum Computation by Contracting Tensor
  Networks</p></title>
  <aug>
    <au><snm>Markov</snm><fnm>IL</fnm></au>
    <au><snm>Shi</snm><fnm>Y</fnm></au>
  </aug>
  <source>SIAM Journal on Computing</source>
  <publisher>Society for Industrial & Applied Mathematics (SIAM)</publisher>
  <pubdate>2008</pubdate>
  <volume>38</volume>
  <issue>3</issue>
  <fpage>963–981</fpage>
  <url>http://dx.doi.org/10.1137/050644756</url>
</bibl>

<bibl id="B11">
  <title><p>Efficient Classical Simulation of Slightly Entangled Quantum
  Computations</p></title>
  <aug>
    <au><snm>Vidal</snm><fnm>G</fnm></au>
  </aug>
  <source>Physical Review Letters</source>
  <publisher>American Physical Society (APS)</publisher>
  <pubdate>2003</pubdate>
  <volume>91</volume>
  <issue>14</issue>
  <url>http://dx.doi.org/10.1103/PhysRevLett.91.147902</url>
</bibl>

<bibl id="B12">
  <title><p>Simulating the Sycamore quantum supremacy circuits</p></title>
  <aug>
    <au><snm>Pan</snm><fnm>F</fnm></au>
    <au><snm>Zhang</snm><fnm>P</fnm></au>
  </aug>
  <pubdate>2021</pubdate>
</bibl>

<bibl id="B13">
  <title><p>Classical variational simulation of the Quantum Approximate
  Optimization Algorithm</p></title>
  <aug>
    <au><snm>Medvidović</snm><fnm>M</fnm></au>
    <au><snm>Carleo</snm><fnm>G</fnm></au>
  </aug>
  <source>npj Quantum Information</source>
  <publisher>Springer Science and Business Media LLC</publisher>
  <pubdate>2021</pubdate>
  <volume>7</volume>
  <issue>1</issue>
  <url>http://dx.doi.org/10.1038/s41534-021-00440-z</url>
</bibl>

<bibl id="B14">
  <title><p>Quantum Computation and Quantum Information: 10th Anniversary
  Edition</p></title>
  <aug>
    <au><snm>Nielsen</snm><fnm>MA</fnm></au>
    <au><snm>Chuang</snm><fnm>IL</fnm></au>
  </aug>
  <publisher>USA: Cambridge University Press</publisher>
  <edition>10</edition>
  <pubdate>2011</pubdate>
</bibl>

<bibl id="B15">
  <title><p>Density Matrix Quantum Circuit Simulation via the BSP Machine on
  Modern GPU Clusters</p></title>
  <aug>
    <au><snm>Li</snm><fnm>A</fnm></au>
    <au><snm>Subasi</snm><fnm>O</fnm></au>
    <au><snm>Yang</snm><fnm>X</fnm></au>
    <au><snm>Krishnamoorthy</snm><fnm>S</fnm></au>
  </aug>
  <source>Proceedings of the International Conference for High Performance
  Computing, Networking, Storage and Analysis</source>
  <pubdate>2020</pubdate>
</bibl>

<bibl id="B16">
  <title><p>Qiskit: An Open-source Framework for Quantum Computing</p></title>
  <aug>
    <au><snm>Aleksandrowicz</snm><fnm>G</fnm></au>
    <au><snm>Alexander</snm><fnm>T</fnm></au>
    <au><snm>Barkoutsos</snm><fnm>P</fnm></au>
    <au><snm>Bello</snm><fnm>L</fnm></au>
    <au><snm>Ben Haim</snm><fnm>Y</fnm></au>
    <au><snm>Bucher</snm><fnm>D</fnm></au>
    <au><snm>Cabrera Hernandez</snm><fnm>FJ</fnm></au>
    <au><snm>Carballo Franquis</snm><fnm>J</fnm></au>
    <au><snm>Chen</snm><fnm>A</fnm></au>
    <au><snm>Chen</snm><fnm>CF</fnm></au>
  </aug>
  <pubdate>2019</pubdate>
</bibl>

<bibl id="B17">
  <title><p>Circ: A Python framework for creating, editing, and invoking Noisy
  Intermediate Scale Quantum (NISQ) circuits</p></title>
  <source>\url{https://github.com/quantumlib/Cirq}</source>
  <pubdate>2021</pubdate>
</bibl>

<bibl id="B18">
  <title><p>Amazon Braket Python SDK: A python SDK for interacting with quantum
  devices on Amazon Braket</p></title>
  <source>\url{https://github.com/aws/amazon-braket-sdk-python}</source>
  <pubdate>2021</pubdate>
</bibl>

<bibl id="B19">
  <title><p>IBM Quantum</p></title>
  <source>\url{https://quantum-computing.ibm.com/}</source>
  <pubdate>2021</pubdate>
</bibl>

<bibl id="B20">
  <title><p>Quantum Computer Datasheet</p></title>
  <aug>
    <au><cnm>Google</cnm></au>
  </aug>
  <source>\url{https://quantumai.google/hardware/datasheet/weber.pdf}</source>
  <pubdate>2021</pubdate>
</bibl>

<bibl id="B21">
  <title><p>Rigetti</p></title>
  <source>\url{https://www.rigetti.com/get-quantum}</source>
  <pubdate>2021</pubdate>
</bibl>

<bibl id="B22">
  <title><p>IonQ</p></title>
  <source>\url{https://ionq.com/}</source>
  <pubdate>2021</pubdate>
</bibl>

<bibl id="B23">
  <title><p>Honeywell</p></title>
  <source>\url{https://www.honeywell.com/us/en/company/quantum}</source>
  <pubdate>2021</pubdate>
</bibl>

<bibl id="B24">
  <title><p>D-Wave: Leap</p></title>
  <source>\url{https://cloud.dwavesys.com/}</source>
  <pubdate>2021</pubdate>
</bibl>

<bibl id="B25">
  <title><p>Benchmarking quantum co-processors in an application-centric,
  hardware-agnostic and scalable way</p></title>
  <aug>
    <au><snm>Martiel</snm><fnm>S</fnm></au>
    <au><snm>Ayral</snm><fnm>T</fnm></au>
    <au><snm>Allouche</snm><fnm>C</fnm></au>
  </aug>
  <pubdate>2021</pubdate>
</bibl>

<bibl id="B26">
  <title><p>Application-Oriented Performance Benchmarks for Quantum
  Computing</p></title>
  <aug>
    <au><snm>Lubinski</snm><fnm>T</fnm></au>
    <au><snm>Johri</snm><fnm>S</fnm></au>
    <au><snm>Varosy</snm><fnm>P</fnm></au>
    <au><snm>Coleman</snm><fnm>J</fnm></au>
    <au><snm>Zhao</snm><fnm>L</fnm></au>
    <au><snm>Necaise</snm><fnm>J</fnm></au>
    <au><snm>Baldwin</snm><fnm>CH</fnm></au>
    <au><snm>Mayer</snm><fnm>K</fnm></au>
    <au><snm>Proctor</snm><fnm>T</fnm></au>
  </aug>
  <pubdate>2021</pubdate>
</bibl>

<bibl id="B27">
  <title><p>The LINPACK benchmark: Past, present, and future</p></title>
  <aug>
    <au><snm>Dongarra</snm><fnm>JJ</fnm></au>
    <au><snm>Luszczek</snm><fnm>P</fnm></au>
    <au><snm>Petitet</snm><fnm>A</fnm></au>
  </aug>
  <pubdate>2002</pubdate>
</bibl>

<bibl id="B28">
  <title><p>The Nas Parallel Benchmarks</p></title>
  <aug>
    <au><snm>Bailey</snm><fnm>D.H.</fnm></au>
    <au><snm>Barszcz</snm><fnm>E.</fnm></au>
    <au><snm>Barton</snm><fnm>J.T.</fnm></au>
    <au><snm>Browning</snm><fnm>D.S.</fnm></au>
    <au><snm>Carter</snm><fnm>R.L.</fnm></au>
    <au><snm>Dagum</snm><fnm>L.</fnm></au>
    <au><snm>Fatoohi</snm><fnm>R.A.</fnm></au>
    <au><snm>Frederickson</snm><fnm>P.O.</fnm></au>
    <au><snm>Lasinski</snm><fnm>T.A.</fnm></au>
    <au><snm>Schreiber</snm><fnm>R.S.</fnm></au>
    <au><snm>Simon</snm><fnm>H.D.</fnm></au>
    <au><snm>Venkatakrishnan</snm><fnm>V.</fnm></au>
    <au><snm>Weeratunga</snm><fnm>S.K.</fnm></au>
  </aug>
  <source>The International Journal of Supercomputing Applications</source>
  <pubdate>1991</pubdate>
  <volume>5</volume>
  <issue>3</issue>
  <fpage>63</fpage>
  <lpage>73</lpage>
  <url>https://doi.org/10.1177/109434209100500306</url>
</bibl>

<bibl id="B29">
  <title><p>ImageNet Large Scale Visual Recognition Challenge</p></title>
  <aug>
    <au><snm>Russakovsky</snm><fnm>O</fnm></au>
    <au><snm>Deng</snm><fnm>J</fnm></au>
    <au><snm>Su</snm><fnm>H</fnm></au>
    <au><snm>Krause</snm><fnm>J</fnm></au>
    <au><snm>Satheesh</snm><fnm>S</fnm></au>
    <au><snm>Ma</snm><fnm>S</fnm></au>
    <au><snm>Huang</snm><fnm>Z</fnm></au>
    <au><snm>Karpathy</snm><fnm>A</fnm></au>
    <au><snm>Khosla</snm><fnm>A</fnm></au>
    <au><snm>Bernstein</snm><fnm>M</fnm></au>
    <au><snm>Berg</snm><fnm>AC</fnm></au>
    <au><snm>Fei Fei</snm><fnm>L</fnm></au>
  </aug>
  <pubdate>2015</pubdate>
</bibl>

<bibl id="B30">
  <title><p>GLUE: A Multi-Task Benchmark and Analysis Platform for Natural
  Language Understanding</p></title>
  <aug>
    <au><snm>Wang</snm><fnm>A</fnm></au>
    <au><snm>Singh</snm><fnm>A</fnm></au>
    <au><snm>Michael</snm><fnm>J</fnm></au>
    <au><snm>Hill</snm><fnm>F</fnm></au>
    <au><snm>Levy</snm><fnm>O</fnm></au>
    <au><snm>Bowman</snm><fnm>SR</fnm></au>
  </aug>
  <pubdate>2019</pubdate>
</bibl>

<bibl id="B31">
  <title><p>MLCommons</p></title>
  <source>\url{https://mlcommons.org/en/}</source>
  <pubdate>2021</pubdate>
</bibl>

<bibl id="B32">
  <title><p>{MLPerf Training Benchmark}</p></title>
  <aug>
    <au><snm>{Mattson}</snm><fnm>P</fnm></au>
    <au><snm>{Cheng}</snm><fnm>C</fnm></au>
    <au><snm>{Coleman}</snm><fnm>C</fnm></au>
    <au><snm>{Diamos}</snm><fnm>G</fnm></au>
    <au><snm>{Micikevicius}</snm><fnm>P</fnm></au>
    <au><snm>{Patterson}</snm><fnm>D</fnm></au>
    <au><snm>{Tang}</snm><fnm>H</fnm></au>
    <au><snm>{Wei}</snm><fnm>GY</fnm></au>
    <au><snm>{Bailis}</snm><fnm>P</fnm></au>
    <au><snm>{Bittorf}</snm><fnm>V</fnm></au>
    <au><snm>{Brooks}</snm><fnm>D</fnm></au>
    <au><snm>{Chen}</snm><fnm>D</fnm></au>
    <au><snm>{Dutta}</snm><fnm>D</fnm></au>
    <au><snm>{Gupta}</snm><fnm>U</fnm></au>
    <au><snm>{Hazelwood}</snm><fnm>K</fnm></au>
    <au><snm>{Hock}</snm><fnm>A</fnm></au>
    <au><snm>{Huang}</snm><fnm>X</fnm></au>
    <au><snm>{Ike}</snm><fnm>A</fnm></au>
    <au><snm>{Jia}</snm><fnm>B</fnm></au>
    <au><snm>{Kang}</snm><fnm>D</fnm></au>
    <au><snm>{Kanter}</snm><fnm>D</fnm></au>
    <au><snm>{Kumar}</snm><fnm>N</fnm></au>
    <au><snm>{Liao}</snm><fnm>J</fnm></au>
    <au><snm>{Ma}</snm><fnm>G</fnm></au>
    <au><snm>{Narayanan}</snm><fnm>D</fnm></au>
    <au><snm>{Oguntebi}</snm><fnm>T</fnm></au>
    <au><snm>{Pekhimenko}</snm><fnm>G</fnm></au>
    <au><snm>{Pentecost}</snm><fnm>L</fnm></au>
    <au><snm>{Janapa Reddi}</snm><fnm>V</fnm></au>
    <au><snm>{Robie}</snm><fnm>T</fnm></au>
    <au><snm>{St. John}</snm><fnm>T</fnm></au>
    <au><snm>{Tabaru}</snm><fnm>T</fnm></au>
    <au><snm>{Wu}</snm><fnm>CJ</fnm></au>
    <au><snm>{Xu}</snm><fnm>L</fnm></au>
    <au><snm>{Yamazaki}</snm><fnm>M</fnm></au>
    <au><snm>{Young}</snm><fnm>C</fnm></au>
    <au><snm>{Zaharia}</snm><fnm>M</fnm></au>
  </aug>
  <source>arXiv e-prints</source>
  <pubdate>2019</pubdate>
  <fpage>arXiv:1910.01500</fpage>
</bibl>

<bibl id="B33">
  <title><p>International SAT Competition</p></title>
  <source>\url{http://www.satcompetition.org/}</source>
  <pubdate>2021</pubdate>
</bibl>

<bibl id="B34">
  <title><p>Shift Scheduling Benchmark Data Sets</p></title>
  <source>\url{http://www.schedulingbenchmarks.org/other.html}</source>
</bibl>

<bibl id="B35">
  <title><p>{Google} cluster-usage traces v3</p></title>
  <aug>
    <au><snm>Wilkes</snm><fnm>J</fnm></au>
  </aug>
  <source>Technical Report</source>
  <publisher>Mountain View, CA, USA</publisher>
  <pubdate>2020</pubdate>
  <note>Posted at
  \url{https://github.com/google/cluster-data/blob/master/ClusterData2019.md}</note>
</bibl>

<bibl id="B36">
  <title><p>{TSPLIB — A Traveling Salesman Problem Library}</p></title>
  <aug>
    <au><snm>Reinelt</snm><fnm>G</fnm></au>
  </aug>
  <source>INFORMS Journal on Computing</source>
  <pubdate>1991</pubdate>
  <volume>3</volume>
  <issue>4</issue>
  <fpage>376</fpage>
  <lpage>384</lpage>
  <url>https://ideas.repec.org/a/inm/orijoc/v3y1991i4p376-384.html</url>
</bibl>

<bibl id="B37">
  <title><p>TSPLIB 95</p></title>
  <aug>
    <au><snm>Reinelt</snm><fnm>G</fnm></au>
  </aug>
  <source>\url{http://comopt.ifi.uni-heidelberg.de/software/TSPLIB95/tsp95.pdf}</source>
  <pubdate>last accessed 2021</pubdate>
</bibl>

<bibl id="B38">
  <title><p>A volumetric framework for quantum computer benchmarks</p></title>
  <aug>
    <au><snm>Blume Kohout</snm><fnm>R</fnm></au>
    <au><snm>Young</snm><fnm>KC</fnm></au>
  </aug>
  <source>{Quantum}</source>
  <publisher>{Verein zur F{\"{o}}rderung des Open Access Publizierens in den
  Quantenwissenschaften}</publisher>
  <pubdate>2020</pubdate>
  <volume>4</volume>
  <fpage>362</fpage>
  <url>https://doi.org/10.22331/q-2020-11-15-362</url>
</bibl>

<bibl id="B39">
  <title><p>Quality, Speed, and Scale: three key attributes to measure the
  performance of near-term quantum computers</p></title>
  <aug>
    <au><snm>Wack</snm><fnm>A</fnm></au>
    <au><snm>Paik</snm><fnm>H</fnm></au>
    <au><snm>Javadi Abhari</snm><fnm>A</fnm></au>
    <au><snm>Jurcevic</snm><fnm>P</fnm></au>
    <au><snm>Faro</snm><fnm>I</fnm></au>
    <au><snm>Gambetta</snm><fnm>JM</fnm></au>
    <au><snm>Johnson</snm><fnm>BR</fnm></au>
  </aug>
  <pubdate>2021</pubdate>
</bibl>

<bibl id="B40">
  <title><p>Benchmarking Quantum Annealing Controls with Portfolio
  Optimization</p></title>
  <aug>
    <au><snm>Grant</snm><fnm>E</fnm></au>
    <au><snm>Humble</snm><fnm>TS</fnm></au>
    <au><snm>Stump</snm><fnm>B</fnm></au>
  </aug>
  <source>Physical Review Applied</source>
  <pubdate>2021</pubdate>
  <volume>15</volume>
  <issue>1</issue>
</bibl>

<bibl id="B41">
  <title><p>The potential of quantum annealing for rapid solution structure
  identification</p></title>
  <aug>
    <au><snm>Pang</snm><fnm>Y</fnm></au>
    <au><snm>Coffrin</snm><fnm>C</fnm></au>
    <au><snm>Lokhov</snm><fnm>AY</fnm></au>
    <au><snm>Vuffray</snm><fnm>M</fnm></au>
  </aug>
  <source>Constraints</source>
  <pubdate>2020</pubdate>
</bibl>

<bibl id="B42">
  <title><p>Readiness of Quantum Optimization Machines for Industrial
  Applications</p></title>
  <aug>
    <au><snm>Perdomo Ortiz</snm><fnm>A</fnm></au>
    <au><snm>Feldman</snm><fnm>A</fnm></au>
    <au><snm>Ozaeta</snm><fnm>A</fnm></au>
    <au><snm>Isakov</snm><fnm>SV</fnm></au>
    <au><snm>Zhu</snm><fnm>Z</fnm></au>
    <au><snm>O'Gorman</snm><fnm>B</fnm></au>
    <au><snm>Katzgraber</snm><fnm>HG</fnm></au>
    <au><snm>Diedrich</snm><fnm>A</fnm></au>
    <au><snm>Neven</snm><fnm>H</fnm></au>
    <au><snm>Kleer</snm><fnm>J</fnm></au>
    <au><snm>Lackey</snm><fnm>B</fnm></au>
    <au><snm>Biswas</snm><fnm>R</fnm></au>
  </aug>
  <source>Phys. Rev. Applied</source>
  <publisher>American Physical Society</publisher>
  <pubdate>2019</pubdate>
  <volume>12</volume>
  <fpage>014004</fpage>
  <url>https://link.aps.org/doi/10.1103/PhysRevApplied.12.014004</url>
</bibl>

<bibl id="B43">
  <title><p>Multi-car paint shop optimization with quantum
  annealing</p></title>
  <aug>
    <au><snm>Yarkoni</snm><fnm>S</fnm></au>
    <au><snm>Alekseyenko</snm><fnm>A</fnm></au>
    <au><snm>Streif</snm><fnm>M</fnm></au>
    <au><snm>Dollen</snm><fnm>DV</fnm></au>
    <au><snm>Neukart</snm><fnm>F</fnm></au>
    <au><snm>Bäck</snm><fnm>T</fnm></au>
  </aug>
  <pubdate>2021</pubdate>
</bibl>

<bibl id="B44">
  <title><p>Benchmarking the quantum approximate optimization
  algorithm</p></title>
  <aug>
    <au><snm>Willsch</snm><fnm>M</fnm></au>
    <au><snm>Willsch</snm><fnm>D</fnm></au>
    <au><snm>Raedt</snm><fnm>H</fnm></au>
    <au><snm>Michielsen</snm><fnm>K</fnm></au>
  </aug>
  <source>Quantum Information Processing</source>
  <pubdate>2020</pubdate>
  <volume>19</volume>
</bibl>

<bibl id="B45">
  <title><p>Quantum Chemistry as a Benchmark for Near-Term Quantum
  Computers</p></title>
  <aug>
    <au><snm>McCaskey</snm><fnm>AJ</fnm></au>
    <au><snm>Parks</snm><fnm>ZP</fnm></au>
    <au><snm>Jakowski</snm><fnm>J</fnm></au>
    <au><snm>Moore</snm><fnm>SV</fnm></au>
    <au><snm>Morris</snm><fnm>T.</fnm></au>
    <au><snm>Humble</snm><fnm>TS</fnm></au>
    <au><snm>Pooser</snm><fnm>RC</fnm></au>
  </aug>
  <pubdate>2019</pubdate>
</bibl>

<bibl id="B46">
  <title><p>An application benchmark for fermionic quantum
  simulations</p></title>
  <aug>
    <au><snm>Dallaire Demers</snm><fnm>PL</fnm></au>
    <au><snm>Stechly</snm><fnm>M</fnm></au>
    <au><snm>Gonthier</snm><fnm>JF</fnm></au>
    <au><snm>Bashige</snm><fnm>NT</fnm></au>
    <au><snm>Romero</snm><fnm>J</fnm></au>
    <au><snm>Cao</snm><fnm>Y</fnm></au>
  </aug>
  <pubdate>2020</pubdate>
</bibl>

<bibl id="B47">
  <title><p>Application-Motivated, Holistic Benchmarking of a Full Quantum
  Computing Stack</p></title>
  <aug>
    <au><snm>Mills</snm><fnm>D</fnm></au>
    <au><snm>Sivarajah</snm><fnm>S</fnm></au>
    <au><snm>Scholten</snm><fnm>TL</fnm></au>
    <au><snm>Duncan</snm><fnm>R</fnm></au>
  </aug>
  <source>Quantum</source>
  <publisher>Verein zur Forderung des Open Access Publizierens in den
  Quantenwissenschaften</publisher>
  <pubdate>2021</pubdate>
  <volume>5</volume>
  <fpage>415</fpage>
  <url>http://dx.doi.org/10.22331/q-2021-03-22-415</url>
</bibl>

<bibl id="B48">
  <title><p>Industry Quantum Applications</p></title>
  <aug>
    <au><snm>Klepsch</snm><fnm>J</fnm></au>
    <au><snm>Kopp</snm><fnm>J</fnm></au>
    <au><snm>Luckow</snm><fnm>A</fnm></au>
    <au><snm>Weiss</snm><fnm>H</fnm></au>
    <au><snm>Standen</snm><fnm>B</fnm></au>
    <au><snm>Vozl</snm><fnm>D</fnm></au>
    <au><snm>Utschig Utschig</snm><fnm>C</fnm></au>
    <au><snm>Streif</snm><fnm>M</fnm></au>
    <au><snm>Strohm</snm><fnm>T</fnm></au>
    <au><snm>Ehm</snm><fnm>H</fnm></au>
    <au><snm>Luber</snm><fnm>S</fnm></au>
    <au><snm>Richter</snm><fnm>J</fnm></au>
    <au><snm>Zeyen</snm><fnm>B</fnm></au>
    <au><snm>Harbach</snm><fnm>P</fnm></au>
    <au><snm>Ehmer</snm><fnm>T</fnm></au>
    <au><snm>Bayerstadler</snm><fnm>A</fnm></au>
    <au><snm>Winter</snm><fnm>F</fnm></au>
    <au><snm>Becquin</snm><fnm>G</fnm></au>
    <au><snm>Polenz</snm><fnm>C</fnm></au>
    <au><snm>Mauerer</snm><fnm>W</fnm></au>
    <au><snm>Niedermeier</snm><fnm>C</fnm></au>
    <au><snm>Gaus</snm><fnm>N</fnm></au>
    <au><snm>Leib</snm><fnm>M</fnm></au>
    <au><snm>Neukart</snm><fnm>F</fnm></au>
    <au><snm>Botter</snm><fnm>T</fnm></au>
    <au><snm>Sepulveda</snm><fnm>J</fnm></au>
    <au><snm>Dombrowski</snm><fnm>E</fnm></au>
    <au><snm>Leonhardt</snm><fnm>T</fnm></au>
    <au><snm>Vornehm</snm><fnm>N</fnm></au>
  </aug>
  <source>\url{https://www.qutac.de/wp-content/uploads/2021/07/QUTAC_Paper.pdf}</source>
  <pubdate>2021</pubdate>
</bibl>

<bibl id="B49">
  <title><p>Computer Systems Performance Evaluation</p></title>
  <aug>
    <au><snm>Ferrari</snm><fnm>D</fnm></au>
  </aug>
  <publisher>Prentice-Hall</publisher>
  <pubdate>1978</pubdate>
  <url>https://books.google.de/books?id=geBQAAAAMAAJ</url>
</bibl>

<bibl id="B50">
  <title><p>The art of computer systems performance analysis - techniques for
  experimental design, measurement, simulation, and modeling</p></title>
  <aug>
    <au><snm>Jain</snm><fnm>R</fnm></au>
  </aug>
  <publisher>Wiley</publisher>
  <series><title><p>Wiley professional computing</p></title></series>
  <pubdate>1991</pubdate>
</bibl>

<bibl id="B51">
  <title><p>Tidy data</p></title>
  <aug>
    <au><snm>Wickham</snm><fnm>H</fnm></au>
    <au><cnm>others</cnm></au>
  </aug>
  <source>Journal of Statistical Software</source>
  <publisher>Foundation for Open Access Statistics</publisher>
  <pubdate>2014</pubdate>
  <volume>59</volume>
  <issue>10</issue>
  <fpage>1</fpage>
  <lpage>-23</lpage>
</bibl>

</refgrp>
} 
\end{document}